\documentclass[twocolumn]{aastex62}

\usepackage{amsmath} 
\usepackage{multirow}
\usepackage{hhline}
\usepackage{xcolor}
\usepackage{rotating}

\defcitealias{Sako2018_SDSS}{S18}
\defcitealias{Heringer2017_DTD}{H17}
\defcitealias{Maoz2012_fig}{M12}
\defcitealias{Dilday2010_SNrate}{D10}

\def\ssnrl{sSNR_L}
\def\vespa{VESPA}
\def\cl{CL}
\def\sfhr{SFHR}
\def\Dcd{\Delta(g-r)}
\def\to{t_{\rm{WD}}}
\def\tc{t_{\rm{c}}}
\def\tone{1\, \rm{Gyr}}
\def\relation{$\Dcd$--$\ssnrl$}
\def\slopes{$s_1$--$s_2$}
\def\As{$A$-$s$}

\def\unity{\rm{yr}^{-1}}
\def\unitV{{\rm dex}(M_\odot ^{-1}\, \rm{yr}^{-1})}

\def\unitS{L_{r,\odot} ^{-1}\, \rm{yr}^{-1}}
\def\PEunit{M_\odot ^{-1}}

\def\CompCLA{$\rm{log}\ A= -12.24 ^{+0.09} _{-0.11}$ $\unitV$}
\def\CompCLs{$s= -1.25 ^{+0.16} _{-0.15}$}
\def\CompVA{$\rm{log}\ A= -12.48 ^{+0.07} _{-0.09}$ $\unitV$}
\def\CompVs{$s= -1.24 ^{+0.17} _{-0.16}$}
\def\PEforty{$PE = 0.004^{+0.002} _{-0.001}\, \PEunit$}

\def\defA{$\rm{log}\ A= -12.15 ^{+0.10} _{-0.13}$ $\unitV$}
\def\defs{$s= -1.34 ^{+0.19} _{-0.17}$}
\def\PEhundred{$PE = 0.004^{+0.002} _{-0.001}\, \PEunit$}
\def\VPEhundred{$PE = 0.002^{+0.001} _{-0.001}\, \PEunit$}
\def\defss{$-1.54 < s_2 < -0.30$}

\def\MCLs{$s= -1.02 ^{+0.15} _{-0.13}$}

\def\MVs{$s= -1.18 ^{+0.13} _{-0.13}$}

\shorttitle{The Delay Times of Type Ia Supernova}
\shortauthors{Heringer et al.}

\begin{document}

\title{The Delay Times of Type Ia Supernova}

\author{E. Heringer}
\affiliation{Department of Astronomy \& Astrophysics, University of Toronto, 50 Saint George Street, Toronto, ON, M5S 3H4, Canada}
\author{C. Pritchet}
\affiliation{Department of Physics and Astronomy, University of Victoria, P.O. Box 1700 STN CSC Victoria, BC V8W 2Y2, Canada}
\author{M. H. van Kerkwijk}
\affiliation{Department of Astronomy \& Astrophysics, University of Toronto, 50 Saint George Street, Toronto, ON, M5S 3H4, Canada}

\begin{abstract}
The delay time distribution of Type Ia supernovae (the time-dependent rate of supernovae resulting from a burst of star formation) has been measured using different techniques and in different environments.  Here, we study in detail the distribution for field galaxies, using the SDSS DR7 Stripe 82 supernova sample.  We improve a technique we introduced earlier, which is based on galaxy color and luminosity, and is insensitive to details of the star formation history, to include the normalization. Assuming a power-law dependence of the supernova rate with time, ${\rm DTD}(t)=A(t/{\rm 1\,Gyr})^{s}$, we find a power-law index \defs\ and a normalization \defA, corresponding to a number of type Ia supernovae integrated over a Hubble time of \PEhundred. We also implement a method used by Maoz and collaborators, which is based on star formation history reconstruction, and find that this gives a consistent result for the slope, but a lower, marginally inconsistent normalization. With our normalization, the distribution for field galaxies is made consistent with that derived for cluster galaxies.  Comparing the inferred distribution with predictions from different evolutionary scenarios for type Ia supernovae, we find that our results are intermediate between the various predictions and do not yet constraint the evolutionary path leading to SNe~Ia. 
\end{abstract}


\newpage

\section{Introduction}
\label{sec:introduction}

Carbon-oxygen white dwarfs (CO WDs) are exceptionally stable, long-lived stars. However, if they are heated or compressed by some external mechanism, or if mass transfer causes their mass to approach the \citet{Chandrasekhar1931_mass} limit, the result can be a thermonuclear runaway, consuming the entire star and leaving no compact remnant behind (for a review, see \citealt{Livio2018_review}).

What triggers the explosion of an inert CO WD as a SN~Ia remains a point of contention. It is currently accepted that a binary system is required, but neither the stellar evolutionary path (sometimes referred as the ``SN~Ia channel''), nor the explosion mechanism have been identified. Unlike core-collapse supernovae, which originate in massive and hence luminous stars, no direct direct imaging of the progenitor systems of SNe~Ia has been obtained. 

In this paper, we attempt to constrain the delay time distribution (DTD), which is defined as the SN~Ia rate for a simple stellar population (SSP) per unit stellar mass and as a function of age. This quantity is relevant because proposed progenitor systems of SN~Ia predict distinct DTDs (e.g. \citealt{Greggio2005_dtd, Toonen2014_BPS}); therefore one may gain insight into the correct channel by constraining the parameters of this distribution.

For instance, in double-degenerate scenarios (DD; e.g. \citealt{Tutokov1981_DD,Webbink1984_DD}), the expected rates are determined primarily by the distribution of orbital separations after a common-envelope phase \citep{Ivanova2013_CE} and by the corresponding timescales for orbital decay due to gravitational radiation, leading to a generic prediction a power--law like DTD with a slope of $-1$. On the other hand, in most single-degenerate scenarios (SD; e.g. \citealt{Whelan1973_SD, Nomoto2000_SD_MSandRG}), the predicted rates are sensitive to a number of parameters, including the mass transfer efficiency, the distribution of binary mass ratios, etc.\ (e.g. \citealt{Greggio2005_dtd}). Still, in many of those scenarios, supernovae are produced only when the white-dwarf companions are relatively massive, which implies DTDs with relative low rates at late times.

It should be noted that many SNe~Ia have peculiar signatures, leading to events being classified into subclasses (e.g. \citealt{Branch2009_subclasses}) which in turn may have distinct channels. For our main analysis, we are interested in subluminous, normal and overluminous events, which form the bulk of SNe~Ia and which may result from a common progenitor system (e.g. \citealt{Nugent1995_sequence,Heringer2017_sequence}, but see also \citealt{Wang2013_V0}, \citealt{Childress2015_Ni} and \citealt{Polin2018_multiple}).

The true form of the DTD may be complicated, but based on generic model predictions (e.g. \citealt{Greggio2005_dtd,Ruiter2011_dtd}), we assume a broken power-law,


\begin{equation}
        \textrm{DTD}(t) \equiv
        \left\{ \begin{array}{ll}
            0 & \textrm{if}\,\,\, t < \to \\
            A\, \left( \frac{t}{\tone} \right)^{s_1} & \textrm{if}\,\,\, \to \leq t \leq \tc \\
            B\, \left( \frac{t}{\tone} \right)^{s_2} & \textrm{if}\,\,\, t \geq \tc.
        \end{array} \right.
\label{eq:DTD} 
\end{equation}

\noindent This equation has three free parameters: a scale factor $A$ (in units of $[\unitV]$) and two slopes, $s_1$ and $s_2$, at young and old ages, respectively. The transition between slopes occurs at a ``cutoff'' time $\tc$, and the rate is continuous in value at $\tc$, such that $B=A\, (\tc / 1\,\rm{Gyr})^{s1-s2}$. The DTD is null prior to the time required for the first WDs to form ($\to$, which we also refer to as the ``onset'' time). Following  \citet{Heringer2017_DTD}[hereafter \citetalias{Heringer2017_DTD}], we typically adopt $\to = 100$\,Myr and $\tc = 1$\,Gyr; these values are based on constraints on the lifetime of the WD or its companion.

Below, we summarize a few of the techniques that have been used to probe the DTD: 

\begin{itemize}
    \item Cluster evolution: Clusters of old galaxies at different redshifts are surveyed for SNe~Ia, thus allowing one to measure the SN~Ia rate at multiple lookback times (e.g. \citealt{Maoz2010_dtd}). Under the assumption that the galaxies can be described as SSPs, one can recover rates as a function of cluster age. This method assumes that the galaxies in the sample formed at nearly the same redshift ($z\sim 3$--$4$, e.g. \citealt{Andreon2016_cluster}), and therefore only the late time DTD is probed ($t \gtrsim 3.2$\ Gyr, \citealt{Friedmann2018_ClusterDTD}). This method also assumes that the observed supernovae did not originate from residual young populations \citep{Schawinski2009_sn}.
   
    \item Remnants: \citet{Maoz2010_powerlaw} used the local star formation rate (SFR) near the location of SN~Ia remnants in the Large and Small Magellanic Cloud to infer rates. The advantage of this technique is that it makes use of resolved stellar populations. However, the data has to be binned in time (owing to the small number of events); in addition, it is difficult to recover many SNe~Ia from old populations with this method.  

    \item Volumetric rates: One can measure volumetric rates up to a fixed redshift and derive a DTD by assuming a cosmic SFH (e.g. \citealt{Graur2011_dtd, Perrett2012_DTD, Frohmaier2019_DTD}).
   
    \item SFH reconstruction (\sfhr): Given a large sample of galaxies that are surveyed over a few years, this method attempts to recover the SFH of each galaxy via spectral fitting. Once the distribution of masses as a function of (binned) ages is obtained, one can statistically infer the slope of the DTD based on which galaxies have hosted SNe~Ia and which have not (e.g. \citealt{Brandt2010_dtd,Maoz2011_fig}). This method is discussed in detail in \S \ref{sec:methods}.
    
    \item Color-luminosity (\cl): \citetalias{Heringer2017_DTD} noted that, when expressed in terms of color and luminosity, specific rates are nearly independent of SFH. This allows one to constrain the DTD parameters from a large sample of galaxies without reconstructing their SFHs. This method has the advantage of not requiring binned data, but has poor sensitivity to rates at early times. This method is also described in detail in \S \ref{sec:methods}.
    
\end{itemize}

Some recent results suggest that the DTD is well-constrained, with a power-law slope close to $s=-1$; this is often interpreted as supportive of the DD channel \citep{Totani2008_powerlaw, Maoz2010_dtd, Graur2011_dtd, MaozandMannucci2012_dtd, Sand2012_MENeaCS_SNsurvey, Graur2013_fig, Graur2014_fig, Rodney2014_DTD}. However, other works have indicated steeper slopes; in particular, rates assessed via the cluster evolution method suggest both a steeper slope and a higher production efficiency\footnote{Number of SN~Ia per unit mass expected during a Hubble time; see \S \ref{subsec:prod_eff}.} (e.g. \citealt{Maoz2017_DTD,Friedmann2018_ClusterDTD}). Also, when applied to a similar sample of galaxies, the \cl\ method indicated a steeper slope than the \sfhr\ method.

This paper has three parts. First, we compare the SFHR and \cl\ methods; D. Maoz has kindly provided us with the data used in \citet{Maoz2012_fig}[hereafter \citetalias{Maoz2012_fig}], thus allowing a direct comparison. Second, we use the \cl\ method to provide an up-to-date characterization of the DTD, including its normalization. Third, we investigate whether the production efficiency of SN~Ia and the slope of the DTD need to be different in field and cluster galaxies.

This paper is organized as follows: in \S \ref{sec:data} we describe the samples, all drawn from the SDSS supernova survey. \S \ref{sec:methods} introduces both the \cl\ and the \sfhr\ methods, and in \S \ref{sec:results} we present the results from both of these approaches. We evaluate some of the systematic uncertainties in the \cl\ method in \S \ref{sec:uncertainties}. Our conclusions are presented in \S \ref{sec:ramifications}, and we summarize our work in \S \ref{sec:summary}. 

\section{Data}
\label{sec:data}

In order to constrain the DTD, it is necessary to construct a well-defined  sample of galaxies for which multiple SNe~Ia are observed. As with \citetalias{Maoz2012_fig} and \citetalias{Heringer2017_DTD}, we use data collected by the SDSS-II Supernova Survey \citep{Frieman2008_SNsurvey}, which covered a $\sim$300 deg$^2$ region of Stripe 82 ($-60^{\circ} \lesssim \alpha \lesssim 60^{\circ}$, $-1\fdg25 \lesssim \delta \lesssim 1\fdg25$) over nine months during 2005-2007.
Specifically, we use the final compilation of the observed SN~Ia and their matched host galaxies of \citet{Sako2018_SDSS}[hereafter \citetalias{Sako2018_SDSS}].

Starting with the sample of galaxies, a variety of data cuts (e.g. in limiting redshift and/or magnitude) are used to optimize both the completeness and purity of the sample. Additional cuts may be adopted to remove spurious or undesired objects, such as stars or QSOs.

Also important is the selection of the SN~Ia sample. This is a non-trivial task because there is a range of reliability for reported events: Some have poorly sampled lightcurves, while others have no or low signal-to-noise (S/N) photometry. Ideally, one wants to maximize the number of SNe~Ia, while making sure that the contamination by other events, such as core collapse SNe, is minimized (e.g. \citealt{Dilday2010_SNrate}).

The \citetalias{Sako2018_SDSS} SNe~Ia come in three flavours: \textit{(i)} those which are confirmed via spectroscopy and whose hosts also have spectra [SNIa], \textit{(ii)} those which are typed photometrically (e.g. \citealt{Niemack2009_photoz}) and whose hosts have spectra [zSNIa], and \textit{(iii)} those which are typed photometrically and whose hosts do not have spectra [pSNIa]. As in \citetalias{Maoz2012_fig} and \citetalias{Heringer2017_DTD}, we only include galaxies that have been observed spectroscopically.

Furthermore, identifying the host galaxy can be complicated. \citetalias{Maoz2012_fig} defines the host as that galaxy whose center is closest in angular separation to the SN~Ia. Alternately, \citet{Sullivan2006_SNrates} (see also \citealt{Gao2012_SNmatching,Sako2018_SDSS}) measure an isophotal radius for all candidate hosts, and find the galaxy with the smallest ratio of angular separation to isophotal radius. In both cases, the redshifts of the SN~Ia and its host must be consistent.

Since we will be comparing our results to \citetalias{Maoz2012_fig} and \citetalias{Heringer2017_DTD}, who have also used SNe~Ia from the SDSS-II Supernova Survey, we next discuss selection criteria and data cuts in these works; in addition, we describe a ``standard sample'' which will be used throughout this paper.

\subsection{M12}
\label{subsec:data_M12}

\citetalias{Maoz2012_fig}'s sample of galaxies is derived from that of \citet{Brandt2010_dtd}, which covers most of Stripe 82, with $-51^{\circ} < \alpha < 59^{\circ}$ and $-1\fdg25 \leq \delta \leq 1\fdg25$, and contains $\sim$83\,000 galaxies with spectroscopic data. Of these, about 77\,000 remained after excluding active galactic nuclei and QSOs. \citetalias{Maoz2012_fig} further trimmed this sample to $\sim$66\,400 galaxies by imposing three conditions: \textit{(i)} A lower limit in velocity to remove possible misidentified stars; \textit{(ii)} An upper limit of $\alpha < 57^{\circ}$ to match the sky region adopted by \citet{Dilday2010_SNrate}[hereafter \citetalias{Dilday2010_SNrate}], from which \citetalias{Maoz2012_fig} retrieved part of their SN~Ia sample. \textit{(iii)} A redshift limit of $z<0.4$ based on the \citetalias{Dilday2010_SNrate} efficiency estimates.

\citetalias{Maoz2012_fig}'s sample of 312 SN~Ia is retrieved from \citetalias{Dilday2010_SNrate} (spectroscopic SNe) and \citet{Sako2011_SNphoto} (photometric SNe), and includes objects with at least one photometric measurement with S/N $> 5$ in the $g$, $r$ and $i$ bands, photometric observations near and after maximum, and light curves well-fitted by a standard template. For host identification, \citetalias{Maoz2012_fig} select host candidates with consistent redshifts and projected physical separations less than 30\,kpc, and, in the case of multiple candidates, select the one with the smallest projected angular separation.  They find 61 matches.  

In the redshift range of interest for this work ($0.01 \leq z \leq 0.2$), there are 47\,728 galaxies and 141 SNIa\footnote{\citetalias{Dilday2010_SNrate} list 140 in their Table~1, but there is one additional without an IAU designation.}, of which 51 have hosts matched by \citetalias{Maoz2012_fig}. 


\subsection{H17}
\label{subsec:data_H17}


In \citetalias{Heringer2017_DTD}, the sample of galaxies contained all objects with spectra for which  $-51^{\circ} \lesssim \alpha \lesssim 59^{\circ}$, $-1\fdg258 < \delta < 1\fdg258$, $0.01 < z < 0.2$ and $14 < r < 17.77$ (after correction for Milky Way extinction). The redshift criterion ensured reliable SN~Ia identification \citep{Dilday2010_SNrate,Sako2018_SDSS}, while the host magnitude cut ensured a nearly complete magnitude-limited sample of galaxies with spectroscopy \citep{Strauss2002_SDSS}. \citetalias{Heringer2017_DTD} imposed further cuts of $g < 22.2$ (after extinction correction) and $g_{\rm{err}},\, r_{\rm{err}} < 0.05$ to limit the number of spurious objects, leaving 20\,707 galaxies.

For the SNe~Ia, \citetalias{Heringer2017_DTD} retrieved 53 hosts from \citet{Gao2012_SNmatching}, which were found using the \citet{Sullivan2006_SNrates} technique of minimizing the ratio of angular separation to isophotal size. An upper limit of 3.8 to this relative distance was enforced, below which contamination was estimated to be small ($< 8\%$). In addition, 3 events observed during the survey engineering time in 2004 were included (we will not use these in the present work, since the survey cadence and duration in 2004 were not adequate for a study of rates; \citealt{Sako2018_SDSS}). All these events are classified as SNIa in the \citetalias{Sako2018_SDSS} terminology.

Because the method \citetalias{Heringer2017_DTD} employed to estimate rates relied on colors, another cut was made by imposing $-0.4 \leq \Delta (g-r) \leq 0.08$, where $\Delta (g-r)$ is the galaxy color relative to the color of the red sequence; this reduced the number of galaxies and hosts to 17\,539 and 52, respectively.

\subsection{Comparison between M12 and H17 samples}
\label{subsec:data_comparison}

Under the assumption of a single power-law DTD, \citetalias{Maoz2012_fig} and \citetalias{Heringer2017_DTD} derived slopes that were not in agreement. It is therefore important to understand how their data samples differed.

\citetalias{Maoz2012_fig}'s sample is much larger than \citetalias{Heringer2017_DTD}'s: 66\,400 vs 17\,539 galaxies. This is mostly because of the different redshift ($0.4$ vs $0.2$) and magnitude limits (no cut vs $14 < r < 17.77$): with the redshift and photometric cuts of H17 and, for completeness, the slightly more restrictive sky region of \citetalias{Maoz2012_fig}, the \citetalias{Maoz2012_fig} and \citetalias{Heringer2017_DTD} samples would contain 19\,329 and 17\,253 galaxies, respectively, with 15\,381 galaxies in common.


Another significant difference is the data source. \citetalias{Maoz2012_fig}'s objects were retrieved from the SpecPhotoAll \textit{Table} database, while \citetalias{Heringer2017_DTD}'s were retrieved from the SpecPhoto \textit{View} database, both for the DR7 data release\footnote{Queries in this database were performed with https://skyserver.sdss.org/CasJobs/default.aspx.}. While the implications of each choice are not obvious, it is likely that the \textit{Table} database includes additional imaging in the Stripe 82 region besides that of the Main Galaxy Sample (\citealt{Strauss2002_SDSS}, see \citealt{Brandt2010_dtd} for more details.).

Yet another difference is the sample of SNe~Ia hosts in each work: in the redshifts of interest, \citetalias{Maoz2012_fig} and \citetalias{Heringer2017_DTD} identified 51 and 52 hosts, respectively.  While these numbers seem consistent, there are only 20 SNe~Ia in common.

The majority of SNIa that are present in \citetalias{Heringer2017_DTD} but not in \citetalias{Maoz2012_fig} are also not present in the SNIa sample of \citetalias{Dilday2010_SNrate} (which \citetalias{Maoz2012_fig} is based on; see above). We are not sure why this is the case. For example, the \citetalias{Dilday2010_SNrate} list misses even the spectroscopically confirmed SNe~Ia SN2006fd and SN2006er, both of which had 15 observed epochs with S/N $>$5. Regarding the SNIa that are present in \citetalias{Maoz2012_fig} but not present in \citetalias{Heringer2017_DTD}, these are all, except 5, explained by the photometric cuts enforced of the galaxy sample in \citetalias{Heringer2017_DTD} and would have otherwise been included.

\subsection{This work}
\label{subsec:data_thiswork}

We now define a ``standard sample'', based on the galaxies used in \citetalias{Maoz2012_fig}, which we will use to derive our most likely DTD parameters and to study systematic uncertainties. We summarize the selection criteria for this dataset below. 

\begin{enumerate}
    \item $-51^{\circ} < \alpha < 57^{\circ}$ and $-1\fdg25 < \delta < 1\fdg25$.
    \item $0.01 < z < 0.2$.
    \item $g_{\rm{err}},\, r_{\rm{err}} < 0.1$
    \item Exclude galaxies with duplicate SDSS DR7 IDs.
    \item Exclude SNe~Ia flagged as peculiar in \citetalias{Sako2018_SDSS}.\footnote{Classified as sn00cx, sn02ci or sn02cx, following the notes given in Table 4 of \citetalias{Sako2018_SDSS}.}
\end{enumerate}

For consistency, the choice of sky region follows that of \citetalias{Maoz2012_fig} and we relax the photometric limits imposed in \citetalias{Heringer2017_DTD}; However, given the considerations in \S \ref{subsec:data_comparison}, we adopt the redshift cuts of \citetalias{Heringer2017_DTD}, while still enforcing an upper limit to photometric errors. The list of both spectroscopically and photometrically typed SNe~Ia and their respective hosts is taken from \citetalias{Sako2018_SDSS}\footnote{The hosts in \citetalias{Sako2018_SDSS} are indicated with a SDSS DR8 ID, which, when possible, we match to a SDSS DR7 by querying under the DR8 context, using the PhotoPrimaryDR7 Table.}. In the redshift range of interest, \citetalias{Sako2018_SDSS}'s list contains a total of 215 SNIa and 70 zSNIa. There are only 3 (5) SNIa in the lists of \citetalias{Maoz2012_fig} \citepalias{Heringer2017_DTD} that are not present in the \citetalias{Sako2018_SDSS} compilation.

\begin{figure}
\epsscale{1.15}
\plotone{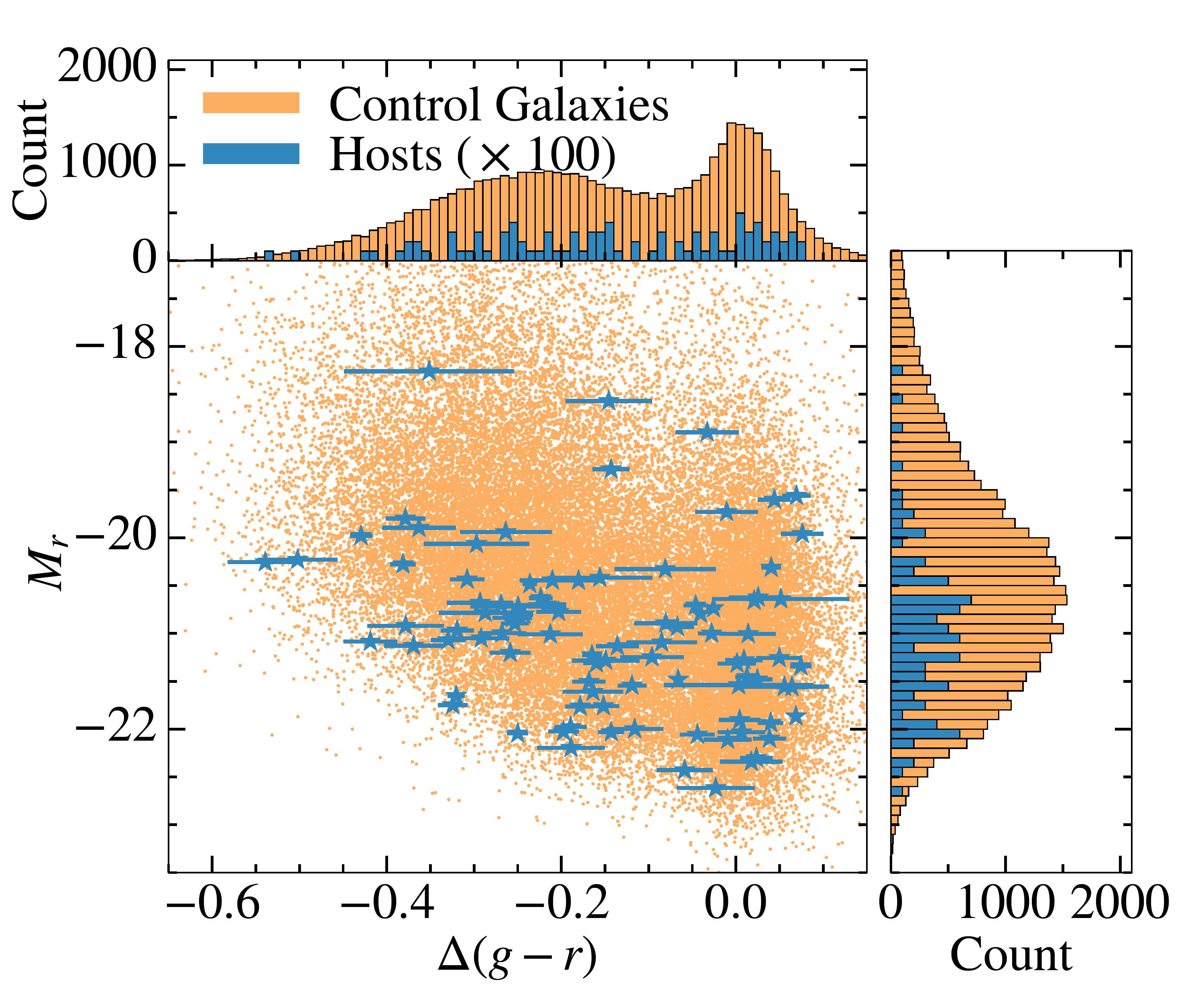}
\caption{The color--magnitude diagram of the galaxies in our standard sample. Colors are measured with respect to the red sequence, for which $\Delta(g - r) = 0$. Host galaxies are shown with blue symbols. Histograms in $\Delta(g - r)$ and in $M_r$ are shown in the upper and right panels, respectively, following the same color code. Host counts are multiplied by 100.}
\label{Fig:CMD}
\end{figure}

In total, this sample contains 43\,895 galaxies and 94 SNe~Ia, out of which 2 SNe~Ia are of peculiar type and therefore not included, leaving 76 SNIa and 16 zSNIa. A color-magnitude diagram of this sample is shown in Fig. \ref{Fig:CMD}.

Before continuing with the above sample, we note that we have tested different combinations of the galaxy samples from \citetalias{Maoz2012_fig} and \citetalias{Heringer2017_DTD} and the hosts from \citetalias{Maoz2012_fig}, \citetalias{Heringer2017_DTD} and \citetalias{Sako2018_SDSS}. As discussed in Appendix \ref{sec:app_dataset}, when the same data cuts are applied, we find DTD slopes that are in agreement within the derived uncertainties, though the normalizations differ.

\section{Methods}
\label{sec:methods}

The SN~Ia rate per unit of mass is, by definition, calculated as

\begin{equation}
r(t) \equiv \textrm{DTD}(t) \ast \textrm{SFR}(t) = \int ^{t} _{\to} \textrm{DTD}(t')\, \textrm{SFR}(t - t')\, dt',
\label{eq:convolution}    
\end{equation}


\noindent where $\to$ is the time at which the first white dwarfs appeared (typically $\to = 100\, \rm{Myr}$, unless we are comparing with \citetalias{Maoz2012_fig}'s results, in which case we adopt their value of $\to = 40\, \rm{Myr}$), and the form of the DTD is given by Eq. \ref{eq:DTD}, with a cutoff time $\tc = 1\, \rm{Gyr}$, unless otherwise stated.

From Eq.~\ref{eq:convolution}, one sees that for a SSP, for which the SFR approaches a delta function, the expected rate will be equivalent to the DTD itself. For the general case, however, the situation is more challenging: the SN rate of each galaxy depends on its SFH.

In this paper, we adopt and compare two methods of analysis: one based on galaxy colours with respect to the red sequence (the \cl\ method), and one based on inferred galaxy masses (the \sfhr\ method), following the prescriptions given in \citetalias{Heringer2017_DTD} and \citetalias{Maoz2012_fig}, respectively. Below we summarize each method, including possible adjustments, and discuss their respective strengths and shortcomings. 

\subsection{The Color-Luminosity (\cl) Method}
\label{subsec:methods_sSNRL}

The \cl\ method of \citetalias{Heringer2017_DTD} uses the fact that the specific supernova rate per unit luminosity ($\ssnrl$) for a galaxy of a given color is quite strongly dependent on the assumed DTD, but nearly independent of the SFH.  In this method, we first use a stellar population code (FSPS; \citealt{Conroy2009_fspsI, Conroy2010_fspsII}) to derive luminosity and color vs. age for some assumed SFH\footnote{The SFH is normalized such that 1 M$_\odot$ is formed over $\sim 14\ \rm{Gyr}$.}, and then compute rates as a function of age via Eq. \ref{eq:convolution}. These results are combined to give a normalized rate per unit luminosity ($\ssnrl$) as a function of color.

The FSPS stellar population is characterized by properties such as the metallicity, initial mass function, etc. These are similar to the assumptions made when reconstructing SFHs and are discussed below (\S~\ref{sec:uncertainties}) as part of the systematic uncertainty budget. To mitigate the impact of some of these unknown parameters, colors are calculated with respect to the red sequence, which is computed as the color of an SSP with age 10\,Gyr. As discussed in \citetalias{Heringer2017_DTD}, we choose to use the $g-r$ color as the locus of the red sequence is clear in these filters and the uncertainties are typically smaller than in $u-r$.


Fig. \ref{Fig:sSNRL} shows predicted \relation\ relations. Two general forms for the SFH are tested: an exponential case, where $\rm{SFR}(t) \propto e^{- t / \tau}$ (left panel) and a delayed exponential case, where $\rm{SFR}(t) \propto t^{-1}\, e^{- t / \tau}$ (right panel), with timescales ranging from $\tau=1$ to $10$\,Gyr.  One sees that the specific rates are nearly independent of the SFH, but are strongly dependent on the DTD. \citetalias{Heringer2017_DTD} showed this to be true also for more complicated SFHs.  The \cl\ method is most sensitive to the slope at later times, $s_2$; to constrain the slope at early ages ($s_1$), one requires very blue galaxies, with $\Dcd \lesssim -0.8$.

For simplicity we ignore (as did \citetalias{Heringer2017_DTD}) the effects of marginalizing our rate predictions over the uncertainty in $\Delta(g-r)$ and $r$-band magnitude. This is justified where the power-law slope of the DTD is shallow, because the supernova rates are nearly independent of $\Dcd$ (see Fig. \ref{Fig:sSNRL}) and the $r$-band error is typically small. On the other hand, for steeper power-law slopes, one sees larger scatter around the \relation\ relation, which likely dominates the uncertainty of these models.

To infer likelihoods for the DTD parameters from the observations, we follow a procedure similar to that of \citetalias{Heringer2017_DTD}, described in detail in Appendix \ref{sec:app_pars}.

\subsection{Improvements to the H17 \cl\ Method }
\label{subsec:methods_sSNRL_mod}

\begin{figure*}
\epsscale{1.12}
\plotone{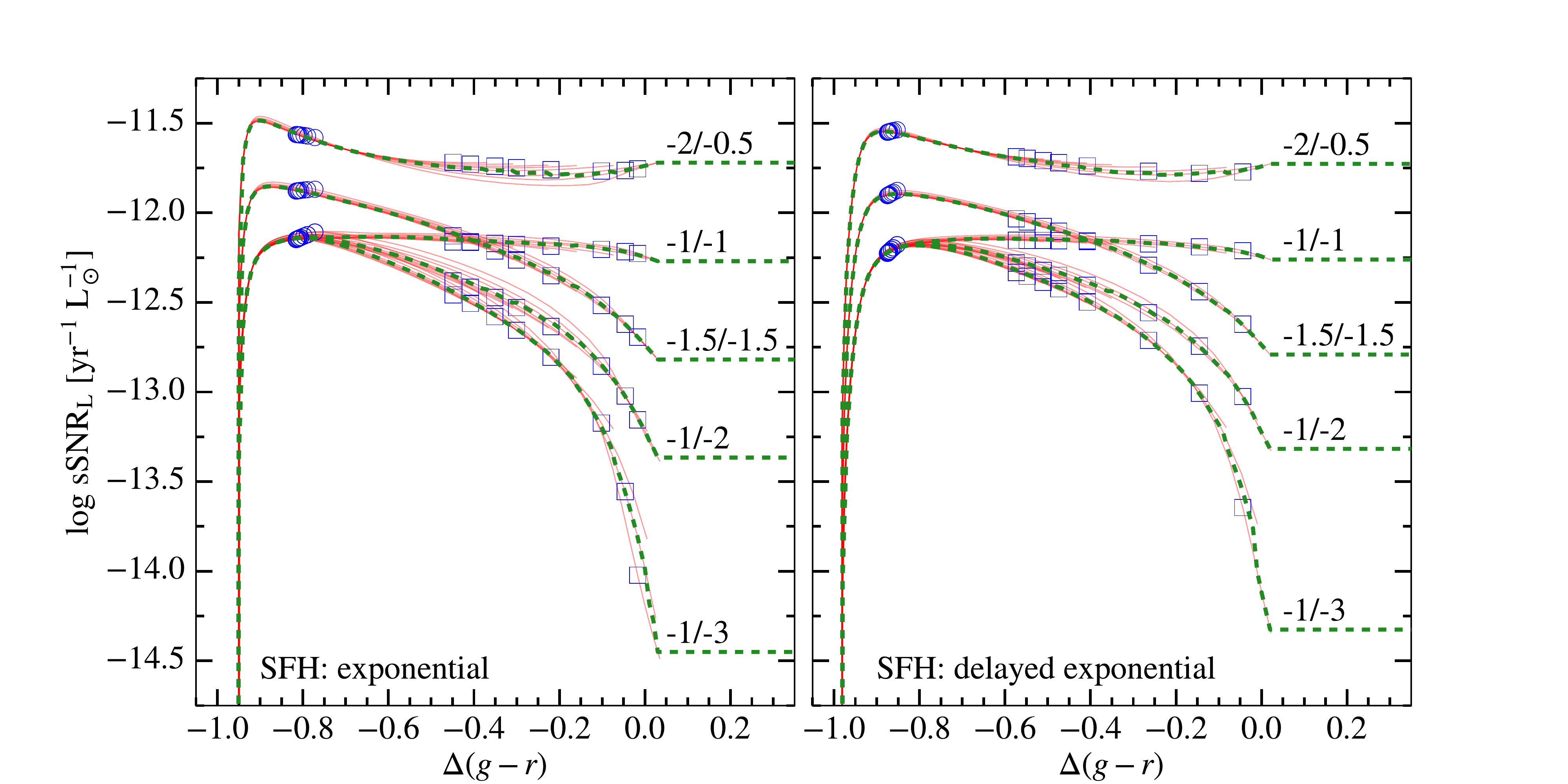}
\caption{SN rate per unit luminosity in the $r$-band as a function of color shift $\Delta(g-r)$ (relative to the red sequence, defined as the color of an SSP with age 10 Gyr). The left panel assumes an exponential SFH of the form of $\rm{SFR} \propto e^{- \frac{t}{\tau}}$, while the right panel assumes $\rm{SFR} \propto t^{-1}\, e^{- \frac{t}{\tau}}$. For a given choice of DTD slopes, the curves (red) correspond to $\tau$, of 1, 1.5, 2, 3, 4, 5, 7, and 10 Gyr. For reference, the loci where $t=1$\ Gyr are marked with empty blue circles and the loci where $t= 10$ Gyr are marked with empty blue squares, for each of the $\tau$ models. The scatter in the predicted rates is small, implying that the SN rate per unit flux is nearly independent of the SFH. The $y$-axis values depend on the normalization of the DTD, which was set to $A=10^{-12}\, \rm{\unitV}$; No vertical shift has been applied to any of the curves. The WD onset time adopted for these calculations is $\to=100$\, Myr.}
\label{Fig:sSNRL}
\end{figure*}

\begin{itemize}
    \item We define our model to be the median of the rates predicted for different SFH's (instead of interpolating the predicted rate per unit luminosity at $10$\, Gyr -- the blue squares in Fig. \ref{Fig:sSNRL}). For the few galaxies that are redder than the color range in our models, we simply adopt the rate at the reddest colour, i.e. $\ssnrl(\Delta > \Delta_{\rm{max}}) = \ssnrl(\Delta_{\rm{max}})$. The final  \relation\ relations are shown as green curves in Fig. \ref{Fig:sSNRL}. With this change, we can use the full range of colours, making it unnecessary to impose cuts on $\Dcd$.
   
    \item For a more straightforward comparison with the \sfhr\ method (see \S \ref{subsec:methods_vespa}), our models assume a Kroupa IMF \citep{Kroupa2007_imf}, rather than the Chabrier IMF \citep{Chabrier2003_imf} used in \citetalias{Heringer2017_DTD}. (This change makes very little difference.)
    
    \item Originally, the \cl\ method was used to probe only the shape of the DTD. Here we extend our models to fit the normalization parameter $A$ in Eq.~\ref{eq:DTD}.

    \item We improve our treatment of $K$-corrections. We use the KCORRECT package \citep{blanton2007_kcorrect}, which requires the input magnitudes to be converted to ``maggies'', a flux unit on a linear scale. In \citetalias{Heringer2017_DTD}, we had made this conversion as if the SDSS photometry were given in magnitudes, when in reality they are given in ``luptitudes'' \citep{Lupton1999_system}; this has been fixed. In addition, following standard routines of the KCORRECT package, we now also perform photometry corrections to the AB system\footnote{These are small: $0.012$ ($0.010$) in the $g$ ($r$)-band.} and add small errors in quadrature in each band.
    
    \item We now take into account effective visibility times, as discussed in Appendix \ref{sec:app_pars} (Eqs. \ref{eq:time_window} and \ref{eq:efficiency}), and we adopt $M_{r,\odot}=4.65$\footnote{Consistent with the FSPS value; see https://python-fsps.readthedocs.io/en/latest/filters/} (e.g. \citealt{Willmer2018_filters}), rather than $M_{r,\odot}=5$.
    
    \item We have also changed the cosmological parameters to make them consistent with those used to compute galaxy masses in \citetalias{Maoz2012_fig} (from WMAP5; \citealt{Komatsu2009_WMAP5}). The new (old) parameters are $H_0=70.5$ ($67$) km s$^{-1}$ Mpc$^{-1}$, $\Omega_\Lambda=0.726$ ($0.68$) and $\Omega_m=0.274$ ($0.32$). We have ensured that these parameters are consistently used in all our calculations.
    
\end{itemize}

Several of the modifications described above are necessary for this method to estimate the scale factor $A$. By comparing stellar masses with those in an independent study \citep{Chang2015_mass}, we argue in Appendix \ref{sec:app_masses} that such a measurement is viable.

We have redone the analysis in \citetalias{Heringer2017_DTD} after incorporating the improvements above, finding that the \citetalias{Heringer2017_DTD} conclusions remain unchanged. The routines used to produce the \relation\ relationships are publicly available\footnote{\url{https://github.com/Heringer-Epson/ssnral}.}. 

\subsection{The SFH Reconstruction (\sfhr) Method}
\label{subsec:methods_vespa}

This method measures the DTD by estimating masses and ages for each galaxy in the dataset. This procedure is similar to that adopted by \citet{Brandt2010_dtd} and by \citet{Maoz2011_fig}, but we focus on the work of \citetalias{Maoz2012_fig}.

Masses in each of a set of age bins (defined in Eq. \ref{eq:mean_ages}) are estimated using the \vespa\ code, by matching a galaxy's spectrum with templates for stellar populations with a \citet{Kroupa2007_imf} IMF and either an exponential or a dual-burst SFH \citep{Tojeiro2007_VESPA}. \citetalias{Maoz2012_fig} uses the masses made available by \citet{Tojeiro2009_VESPA}.

\citetalias{Maoz2012_fig} combined the VESPA masses in three age bins: 0--0.42\,Gyr, 0.42--2.4\,Gyr and $>\!2.4$\,Gyr. They also scaled all masses by a factor of 0.55 to account for light outside of SDSS fibers, etc., verifying that their final masses are consistent with other, independent estimates.

For each galaxy $i$ in their sample, \citetalias{Maoz2012_fig} assign an expected supernova rate summed over the three age bins,
\begin{equation}
r_i \approx m_{i,1} \psi_1 + m_{i,2} \psi_2 +  m_{i,3} \psi_3,
\label{eq:M12_rates}    
\end{equation}
where the numeric indices denote the three time bins. The rates $\psi$ are estimated using the same likelihood calculation as for the CL method, summarized in appendix \ref{sec:app_pars}, except that in Eqs. \ref{eq:likelihood1}--\ref{eq:likelihood7}, the likelihood is kept as a function of $\psi_{1-3}$. In order to determine the slope of the DTD, \citetalias{Maoz2012_fig} fitted a power-law function to the most likely binned rates~$\psi$. 



Following \citetalias{Maoz2012_fig}'s analysis\footnote{Eq.~7 of \citetalias{Maoz2012_fig} contains a typo: the factor $t_i^2$ should be $(\epsilon t_i)^2$.}  we derived maximum likelihood rates that agreed with their Table 2 to within $1\%$. Fitting a DTD slope to these rates yielded slightly different results: $s=-1.23 \pm 0.19$, rather than $s=-1.12 \pm 0.08$. In any case, we show below (in \S \ref{subsec:methods_mod}) that one can assess the DTD most likely parameters directly from the Bayesian analysis, thus avoiding the additional step of fitting a power-law to rates $\psi$.

\subsection{Comments on the \sfhr\ Method}
\label{subsec:methods_comments}

While the rate recovery method is reliable, one needs to make important assumptions in order to derive DTD parameters from those:   

\textit{i)} The rate in each bin was assumed to be representative of a single age, which was taken to be the linear mean age of the bin (Fig. 1 of \citetalias{Maoz2012_fig}).

\textit{ii)} The observed rates were approximated as a binned mass multiplied by the intrinsic DTD rate (see Eq. \ref{eq:M12_rates}). It is difficult to quantify how much this approximation affects the DTD slope, given that it bypasses the convolution in Eq. \ref{eq:convolution}, and given that only three age bins were used.

\textit{iii)} VESPA masses in the youngest bin are overestimated, because they include stars in the age range $0$--$70$\, Myr which cannot be distinguished from later ages ($70$--$420$\, Myr).

\textit{iv)} The \sfhr\ method uses only 3 age bins, and hence depends strongly on the assumed onset time of SN~Ia, $\to$, which is poorly known. Note that \cl\ method is also sensitive to this variable, and its effects are explored in \S \ref{sec:uncertainties}.

\begin{figure}[h]
\epsscale{1.15}
\plotone{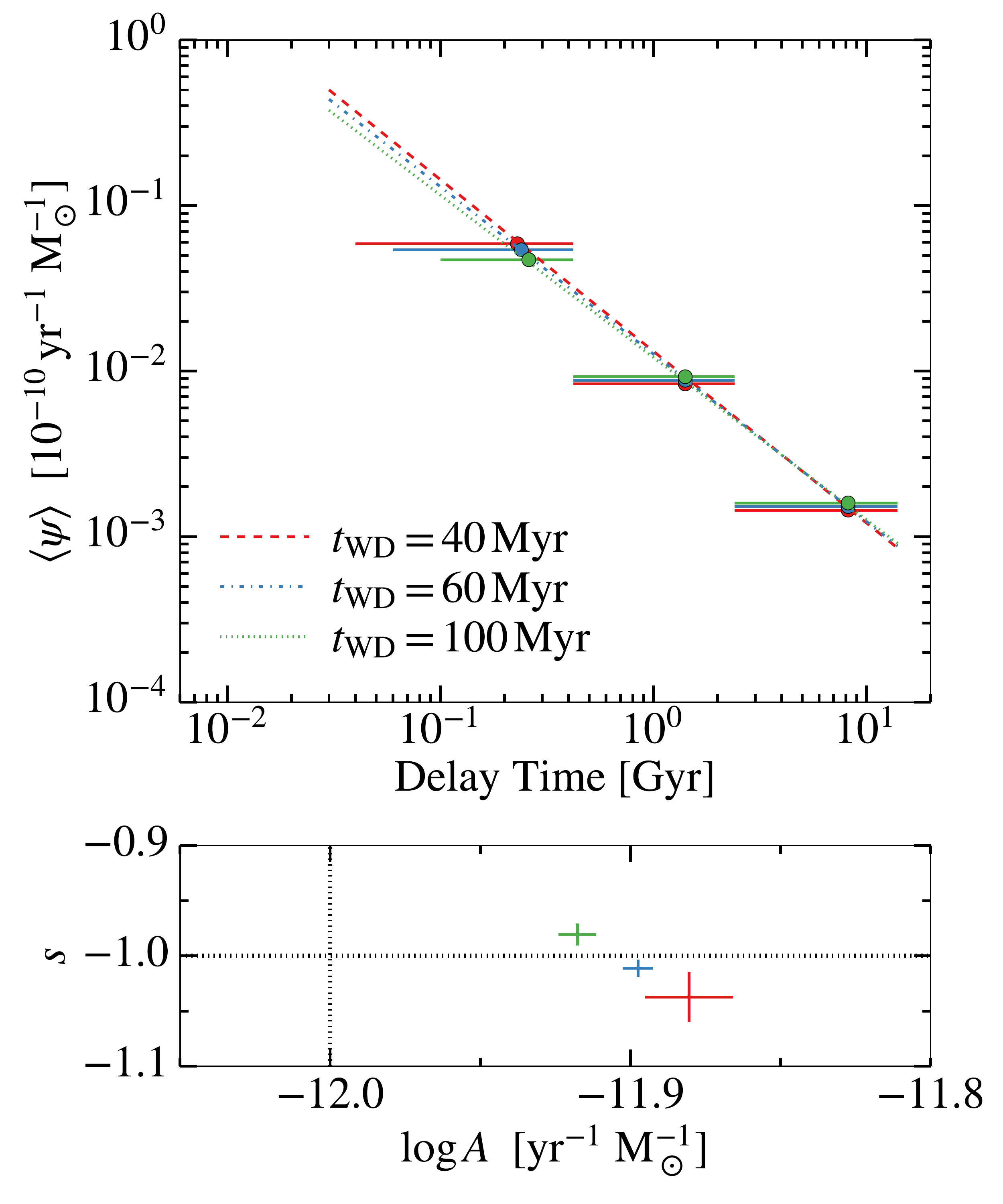}
\caption{\sfhr\ fitting method test, where mock rates are computed via Eq. \ref{eq:approx_expanded} from a known DTD with power law $s=s_1=s_2=-1$. \textit{Top:} A power-law is fitted to the binned rates. Different colors correspond to three choices of the onset time: $\to=40$, $60$ and $100$\, Myr. \textit{Bottom:} The recovered slopes and scale factors, following the same color scheme as the panel above. The input parameters are shown as dotted lines. The recovered slopes and scale factors disagree with the input. Displayed rates have been offset for clarity.}
\label{Fig:test_M12}
\end{figure}

To illustrate the point that fitting a power-law to binned rates may not be an optimal way to recover DTD parameters, and that the the fitted power-law depends on the assumed onset time, we calculate mock mean rates as


\begin{equation}
\langle \psi_j \rangle \approx \langle \rm{DTD}_{j} \rangle = \frac{1}{t_{\rm{f,j}} - t_{\rm{i,j}}}\, \int _{t_{\rm{i,j}}} ^{t_{\rm{f,j}}} \rm{DTD}(t|A,s_1,s_2) dt,
\label{eq:component_rate}    
\end{equation}

\noindent from a known DTD with an assumed slope $s=-1$ and a scale factor $A = 10^{-12}$ $\unitV$; $t_{i,j}$ and $t_{f,j}$ are the initial and final ages of the $j$-th bin. These rates are then fitted by a power-law, as shown in the top panel of Fig. \ref{Fig:test_M12}, for three choices of $\to$ (color coded). The bottom panel shows the recovered slope and scale factor. 

We find that the recovered slope is sensitive to the choice of $\to$; the scale factor is over-predicted in all cases. We also checked that better agreement would be achieved with a larger number of bins. This exercise is not entirely equivalent to \citetalias{Maoz2012_fig}'s procedure, since here the mean DTD rates have been taken directly from the input DTD, without computing binned masses (which would add to the uncertainties). Therefore we will not use this fitting procedure.

\subsection{Adjustments to the \sfhr\ Method}
\label{subsec:methods_mod}

To avoid some of the problems associated with fitting a power law to recovered rates, we compute likelihoods which depend directly on the DTD parameters, by expressing rates as $r_i(t_i|A,s_1,s_2)$ rather than $r_i(t|\psi_{i,1},\psi_{i,2},\psi_{i,3})$. It is still necessary to use the approximation that the expected rates in each galaxy depend on binned quantities. Furthermore, we still need to assign an age to each bin; for consistency with \citetalias{Maoz2012_fig}, we will approximate this as the linear mean in each bin, $\bar{t_j}$.  In summary, we replace Eq.~\ref{eq:M12_rates} with 
\begin{equation}
r_i \approx \sum_{j=1,2,3} m_{i,j} \times \psi_j(\bar{t_j}|A,s_1,s_2).
\label{eq:approx_rates}    
\end{equation}

Note that this formulation allows us to fit a broken power law ($s_1\neq{}s_2$), but the rates in the intermediate age bin, for which $0.42 < t < 2.4$\,Gyr, become awkward if the cutoff time falls within it, as is the case for our default choice of $\tc = 1\,$Gyr.  For the comparison below (\S \ref{subsec:method_comp}), we simply try cut-off times at the start and end of the intermediate bin.

The bin age limits and mean ages and an expanded form of Eq.~\ref{eq:approx_rates} are given in appendix \ref{sec:app_rates}; Likelihoods are computed following the description given in appendix \ref{sec:app_pars}. 

\subsection{Method Comparison on Mock Data}
\label{subsec:methods_mock}

We used FSPS to simulate the color and luminosity of galaxies according to a set of random input parameters, which consists of an SFH timescale, a formed stellar mass, and a galaxy age. With a known DTD slope and scale factor, we used Eq. \ref{eq:convolution} to compute the true rate for each galaxy, based on which a random sample of host galaxies was drawn. We computed the expected rates for each galaxy and confidence contours for the DTD parameters using both the \cl\ and \sfhr\ methods, finding good agreement with the true values, noting that some of the approximations in the \sfhr\ method (see \S \ref{subsec:methods_mod}) cause the rates to be underestimated by $\sim 10\%$ and the derived scale factor to be overestimated at a similar level. This suggests that discrepancies for the best DTD parameters when these methods are applied to real data are likely due to possible systematic errors in either or both techniques, such as the mass estimation in the \sfhr\ method or by having colors computed relative to the red-sequence in the \cl\ method.

\section{Results}
\label{sec:results}

\subsection{Method Comparison}
\label{subsec:method_comp}

\begin{figure*}
\epsscale{1.15}
\plotone{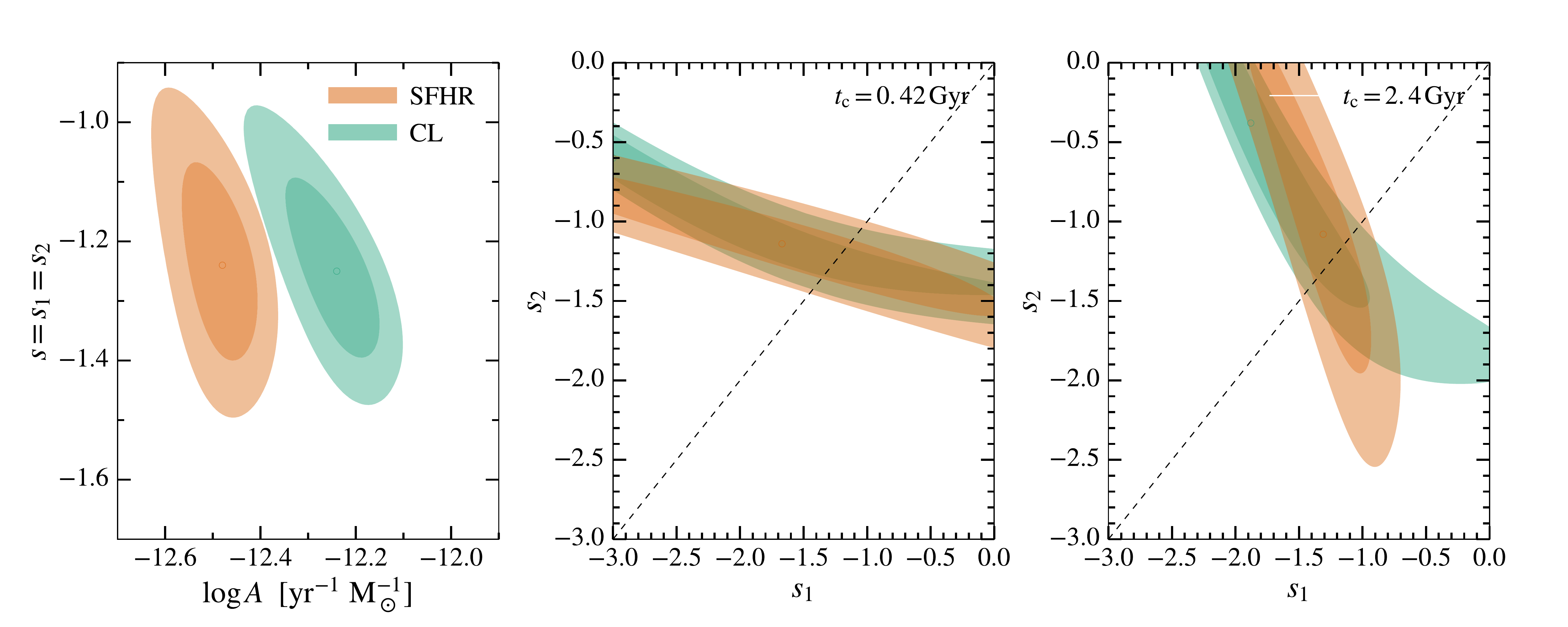}
\caption{Likelihood contours derived from the \cl\ (green) and \sfhr\ (orange) methods. 68\% and 95\% contour regions are shown as dark and light shades, and the location of the most likely parameters are marked with circles. \textit{Left:} \As\ parameter space, where the slope of the DTD is assumed to be continuous ($s=s_1=s_2$). \textit{Middle} and \textit{Right:} \slopes\ parameter space, where the scale factor $A$ is that which maximizes likelihood for each combination of slopes, so that the predicted number of SNe matches the observations. In the middle (right) panel, the cutoff age is $0.42$ ($2.4$)\,Gyr, such that the intermediate age bin in \citetalias{Maoz2012_fig} is described by the DTD slope at late (early) times. We use our standard dataset (see \S \ref{subsec:data_thiswork}) and, following \citetalias{Maoz2012_fig}, we use $\to=40$\, Myr. The left panel indicates that both methods produce consistent slopes, while the scale factor disagrees at the $>2\sigma$ level.}
\label{Fig:method_comp}
\end{figure*}

As a validation exercise, we first compare the likelihoods derived for the modified \sfhr\ and \cl\ methods. For this we use our ``standard'' sample, while adopting $\to=40$\, Myr, as in \citetalias{Maoz2012_fig}.

We note that since we are using a slightly different sample, the fitted DTD parameters will not be identical to the \citetalias{Maoz2012_fig} and \citetalias{Heringer2017_DTD} results. We also caution that the only parameter that can be calculated and compared using the original methods of \citetalias{Maoz2012_fig} and \citetalias{Heringer2017_DTD} is the continuous DTD slope $s=s_1=s_2$. The calculation of the scale factor $A$ and the \slopes\ likelihoods\footnote{As shown in Appendix \ref{sec:app_pars}, one does not need to marginalize the \slopes\ likelihood over the scale factor because the integration over $A$ does not depend on the choice of slopes.} arise from adjustments discussed in \S 3. For the latter calculation, we try cutoff ages at the start and end of the intermediate bin (see \S \ref{subsec:methods_mod}), while the standard case for which $\tc=1$\,Gyr is discussed below.

The left panel of Fig. \ref{Fig:method_comp} shows the \As\ parameter space, calculated via the \cl\ method (green) and the \sfhr\ method (orange). The most likely parameters are retrieved directly from the posterior distribution using Eq. \ref{eq:component_rate}. If a continuous slope is assumed, then \CompCLs\ (\CompVs) when applying the \cl\ (\sfhr) method; these slopes are in excellent agreement. The scale factors however differ at the $>2\sigma$ level:  \CompCLA\ for the \cl\ method and \CompVA\ for the \sfhr\ method. These results indicate a discrepancy by a factor of $\sim 2$ for the most likely DTD normalization; we discuss possible sources for this difference in \S \ref{sec:ramifications}.

The \slopes\ parameter space is shown in the middle and right panels of Fig. \ref{Fig:method_comp}, according to the choice of cutoff time. One can see that neither method is capable of constraining well the early slope if $\tc=0.42$\,Gyr, because of the small fraction of blue galaxies. At the $68\%$ confidence level, we find $-1.46 < s_2 < -0.45$ ($-1.60 < s_2 < -0.72$) for the \cl\ (\sfhr) method. If $\tc=2.4$\,Gyr, then, for both methods, the early slope becomes better constrained, while a larger range of $s_2$ is acceptable. We find $-2.23 < s_1 < -0.94$ and $s_2 > -1.54$ ($-1.90 < s_1 < -0.93$ and $s_2 > -1.96$) for the \cl\ (\sfhr) method.

Overall, we conclude that the two methods give consistent information on the shape of the DTD, but not on its normalization.

\subsection{Results From the Default Sample}
\label{subsec:res_default}

\begin{figure*}
\epsscale{1.15}
\plotone{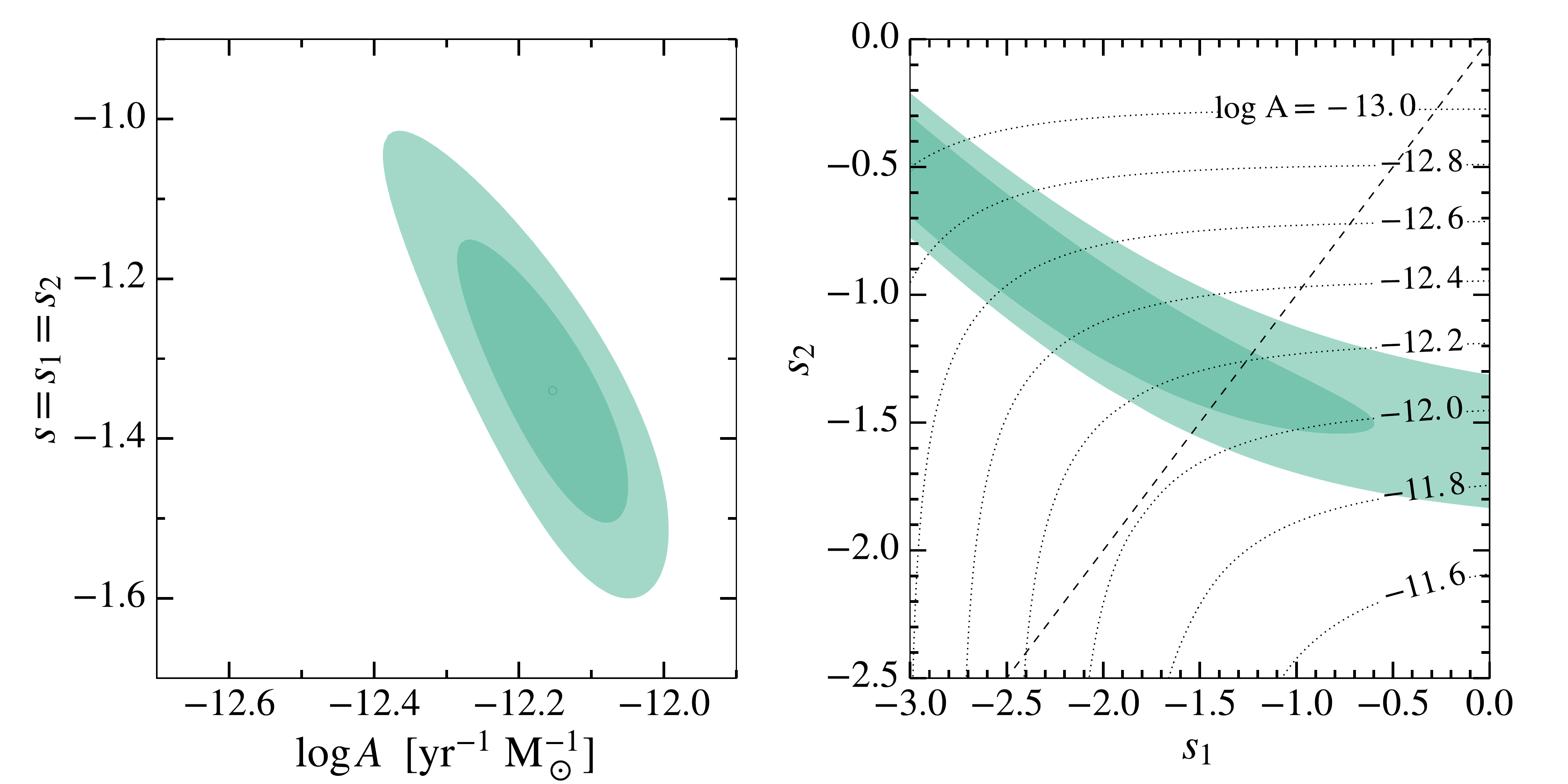}
\caption{Confidence contours for the \cl\ method when applied to our standard dataset. 68\% and 95\% contour regions are shown as dark and light shades. \textit{Left:} ($A,s$) parameter space, where the slope of the DTD is assumed to be continuous ($s=s_1=s_2$). \textit{Right:} ($s_1,s_2$) parameter space; the scale factor is varied so that the predicted number of SNe matches observations. Curves of $\rm{log}\, A$ are shown for reference (dotted lines). A continuous slope $s_1=s_2$ is marked with the dashed line. Following \citetalias{Heringer2017_DTD}, we have adopted $\to=100\ \rm{Myr}$ for these calculations. This sample contains 50 hosts and 17\,253 galaxies.}
\label{Fig:results}
\end{figure*}

We use our ``standard'' sample (defined in \S \ref{subsec:data_thiswork}) to characterize the most likely DTD parameters using the \cl\ method. The confidence contours are shown in Fig.~\ref{Fig:results} differ somewhat from those in Fig.~\ref{Fig:method_comp} because we use our preferred $\to=100$\,Myr.

Assuming a continuous slope ($s=s_1=s_2$; left panel), we find \defA\ and \defs. If instead the slopes at early and late times are allowed to vary, then $s_1$ is poorly constrained due to the small number of SNe~Ia in blue galaxies. As a consequence, $s_2$ will also span a large range and, at the $68\%$ confidence level, we find $s_1 < -0.6$ and \defss. Note that the scale factors (dotted lines) are not the same across this parameter space, but vary such that the predicted number of SNe~Ia matches the observed one.  

\subsection{Production Efficiency}
\label{subsec:prod_eff}

The production efficiency is the number of SN~Ia per unit mass expected from a burst of star formation over a Hubble time (e.g. \citealt{Friedmann2018_ClusterDTD}). It is defined as

\begin{equation}
PE \equiv \int_{\to} ^{13.7\, \rm{Gyr}} \rm{DTD}(t)\, dt.
\label{eq:prod_eff}    
\end{equation}

\noindent This quantity is more physical than either $A$ or $s$, is less sensitive to correlations between $s$ and $A$, and is better for comparing DTDs with different shapes.

By applying \cl\ (\sfhr) method to our default sample under the assumption of a continuous slope $s$, we find \PEhundred\ (\VPEhundred), noting these values do not change appreciably if $\to = 40\, \rm{Myr}$ or $\to = 100\, \rm{Myr}$. We defer a discussion of these values to \S\ref{sec:ramifications}.


\section{Model uncertainties}
\label{sec:uncertainties}

\begin{figure}
\epsscale{1.15}
\plotone{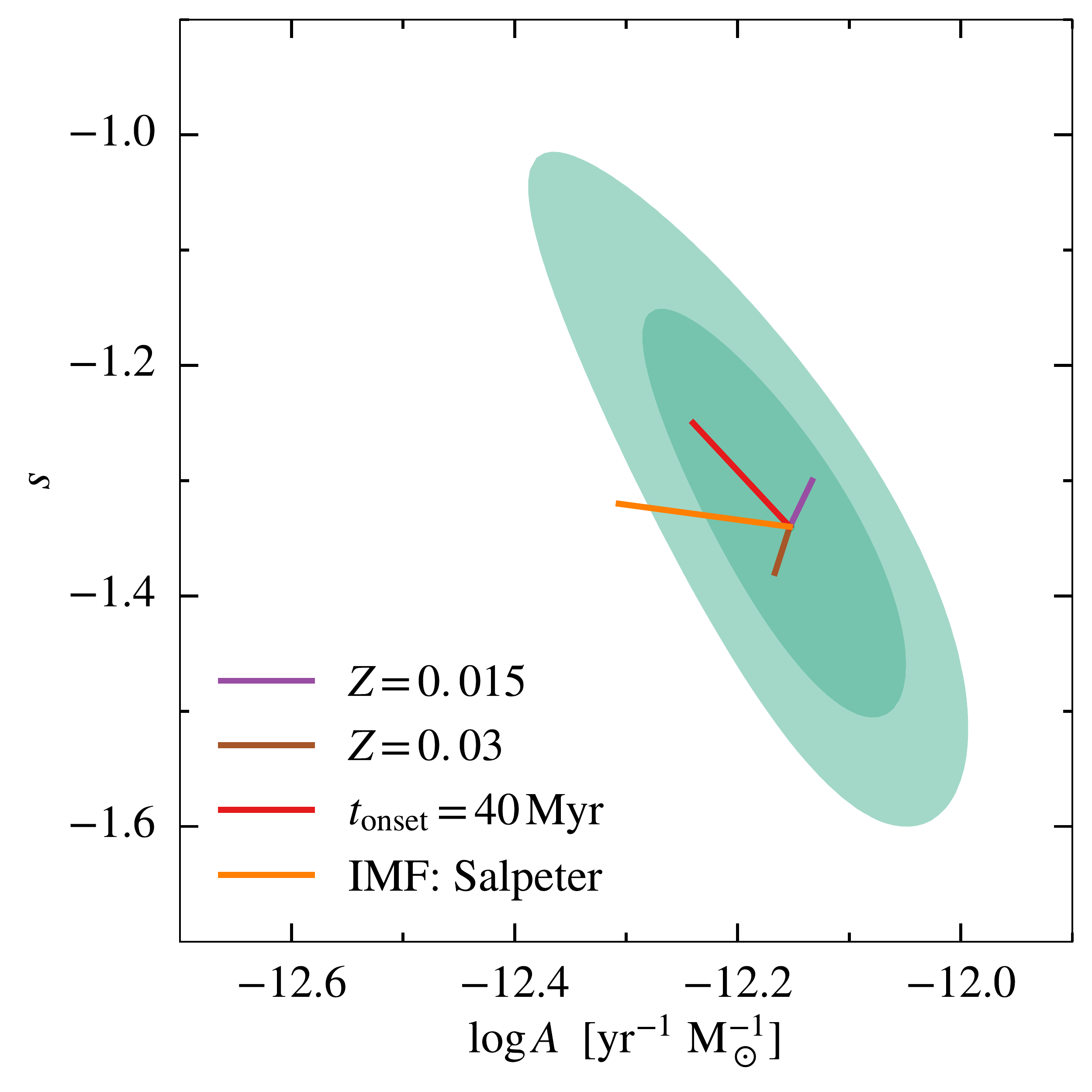}
\caption{Study of systematic uncertainties. The gray confidence contours are derived using the \cl\ method with the fiducial model parameters (\S \ref{sec:uncertainties}). The color coded lines indicate how the position of the DTD fitted parameters changes if the model is changed. Only the model parameters that lead to appreciable changes are shown. These consist of adopting a  \citet{Salpeter1955_imf} with a lower limit of $0.08\,M_\odot$ (instead of a \citealt{Chabrier2003_imf}) IMF, sub and super-solar metallicity ($Z=0.015$ and $Z=0.30$ rather than solar $Z=0.019$) and a shorter time for the formation of the first WDs ($\to\ = 40$\, Myr rather than $\to\ = 100$\,Myr). Qualitatively, one can see that systematic uncertainties are $\sim 0.1$ for the slope $s$, and $0.2\,\unitV$\ for $\rm{log}\,A$.}
\label{Fig:uncertainties}
\end{figure}

In this section we examine how different model assumptions might impact the likelihoods for our standard sample. We use the \cl\ method and focus on the \As\ parameter space, as its likelihood is much better constrained. The ``fiducial'' model parameters include: a Kroupa IMF \citep{Kroupa2001_imf}, an exponential SFH, solar metallicity ($Z=0.019$), no blue horizontal branch stars ($f_{BHB}=0$), the BASEL spectral library \citep{Lejeune1997_BaSelI,Lejeune1998_BaSelII,Westera2002_BaSelIII}, the PADOVA isochrone library \citep{Marigo2007_Padova,Marigo2008_Padova}, an evolutionary factor $Q=1.6$ \citep{Wyder2007_cmd}, and a white dwarf onset time $\to\ = 100$\, Myr.  

We perform several tests, changing the standard model parameters above one at a time. The changes for these tests include: a Chabrier \citep{Chabrier2003_imf} or Salpeter \citep{Salpeter1955_imf} IMF, a delayed-exponential SFH, a metallicity of $Z=0.015$ or $Z=0.03$, $f_{BHB}=0.2$, the MILES spectral library \citep{Sanchez2006_MILES}, the MIST isochrone library \citep{Dotter2016_MIST}, $Q=0$, and $\to\ = 40$\, Myr.

For reference, in Fig. \ref{Fig:uncertainties} we show the confidence contours obtained for the fiducial model parameters. This confidence contour is similar in shape and size for all of our tests, and so the figure shows only the changes in the most likely value of the fit for each test. Furthermore, for clarity we show only those tests which result in a significant change in slope or scale factor ($\Delta s \geq 0.03$ or $\Delta \rm{log}\ A \geq 0.05$). In general we find from the tests that the systematic uncertainies are  $\Delta s \sim 0.1$ $\Delta (\rm{log}\ A) \sim 0.2$. In more detail:

Relative to a Chabrier or Kroupa IMF, the Salpeter IMF has more low-mass stars (bottom-heavy distribution). Other works have indicated that galaxy masses derived assuming a Salpeter IMF tend to be roughly 1.4 times larger than assuming a Chabrier or Kroupa IMF \citep{Cunha2008_MAGPHYS,Maoz2012_fig}. This explains the lower scale factor (orange segment) seen in Fig.~\ref{Fig:uncertainties}.

Changing the metallicity has little effect on the shape of the \relation\ relation shown in Fig. \ref{Fig:sSNRL}, but can lead to a small offset $\Delta(\rm{log}\ \ssnrl) \sim 0.1$ in the rate at $\Dcd \sim 0$. For instance, lower metallicities will shift the confidence contours upwards, such that these lower rates are compensated for by slightly shallower slopes (see the -1/-1 and -1.5/-1.5 cases in Fig.~\ref{Fig:sSNRL}). Also, since the cumulative rates become smaller, a slightly higher scale factor will be preferred. This exercise suggests that the \cl\ method is sensitive to the precise $\ssnrl$ for colors redder than the RS; the assumption of a constant specific rate in that regime may be an oversimplification.

The effect of adopting a shorter $\to$ is that $\ssnrl$ rates become higher for bluer galaxies, but are unchanged for redder objects. As a result, the cumulative rates will be higher, thus leading to a smaller scale factor $A$. Moreover, a shallower slope is preferred because it would boost the relative rates expected for red galaxies, thus compensating for the change in rate for the blue galaxies due to a shorter $\to$.

\subsection{Dust}
\label{subsec:dust}


Our stellar population models are calculated using the FSPS package \citep{Conroy2009_fspsI, Conroy2010_fspsII}, for which different choices of dust parameters are allowed. As in \citetalias{Heringer2017_DTD}, we adopt the default mode for dust absorption (\texttt{dust\_type=0}), which follows the two-component model of \citet{Charlot2000_dust}. This model assumes that all stars are affected by an uniform dust screen and, in addition, young stars also suffer from a power-law attenuation curve. This dust treatment is similar to that adopted in \vespa\ (see \S 2.2.2. of \citealt{Tojeiro2009_VESPA}). We performed additional tests with different choices of attenuation curves, finding that the most likely slope and scale factor did not change appreciably.

\section{Discussion}
\label{sec:ramifications}

We have demonstrated that both the \cl\ and the \sfhr\ methods yield approximately the same power-law slope, $s\simeq-1.35$ for the DTD if no break is assumed. This value is inconsistent with that reported by both \citetalias{Maoz2012_fig} ($s\sim -1.1$) and \citetalias{Heringer2017_DTD} ($s\sim -1.5$), mainly because of the difference between adopted samples (see \S \ref{subsec:data_comparison}). Under the assumption of $\to = 40$\,Myr, if we relax the photometric and redshift limits so that the  \cl\ (\sfhr) method is applied to the entire sample of \citetalias{Maoz2012_fig}, we obtain \MCLs\ (\MVs).

The scale factors derived from both methods are marginally inconsistent, at the $>2\sigma$ level, with the \sfhr\ method suggesting a lower normalization, by a factor of $\sim$2. Both methods are applied to the same sample and thus differences in the data cannot explain this discrepancy. In Appendix~\ref{sec:app_masses} we find that, for a subset of red-sequence galaxies, the mass scaling based on FSPS is consistent with independent source. \citetalias{Maoz2012_fig} also finds the VESPA masses to be consistent with another independent source. In \S~\ref{subsec:methods_mock}, we have verified that both methods are able to accurately recover the DTD parameters, which suggests that the scale factor discrepancy must arise from systematic uncertainties in either or both methods. For instance, VESPA masses are based on magnitudes computed by integrating the SDSS spectra over multiple bands, while the \cl\ method uses photometric magnitudes solely from the $r$-band. Also, due to the nature of the \cl\ method, we can only verify that the masses from galaxies in the red-sequence are consistent, but it is possible that this agreement does not hold for bluer galaxies. 

The slope derived by \citetalias{Maoz2012_fig} is in good agreement with those estimated in volumetric surveys \citep{Perrett2012_DTD,Frohmaier2019_DTD}, which yielded $s\sim-1$. However, we show that when the \sfhr\ method is applied to a more conservative sample (i.e. $z<0.2$), a steeper slope is recovered. Moreover the production efficiency are not in agreement: \citet{Perrett2012_DTD} obtained ($4.4\pm0.2$)--($5.2\pm0.2$)\,$10^{-4}\, \PEunit$, while we obtain value roughly a factor of 10 (4) larger using the \cl\ (\sfhr) method. It is intriguing to us that the cluster and volumetric measurement of the DTD could yield such discrepant production efficiencies, while the \sfhr\ and \cl\ methods seemingly recover intermediate values.

\subsection{DTD Dependency on Environment}
\label{subsec:environment}

We find that our results are in marginal agreement with recent inferences of both the slope and the scale factor of the DTD in cluster galaxies.   In particular, \citet{Friedmann2018_ClusterDTD} obtained $s=-1.45^{+0.51} _{-0.38}$ and $\mathrm{log}\, A = -12.00^{+0.34} _{-0.52} \unitV$, with a production efficiency of $0.009 ^{+0.029} _{-0.007}\, \PEunit$ (see their Table 7\footnote{For fairness of comparison, we use the result that assumes no knowledge of the cluster iron content, and use the values derived using a single redshift bin; the values derived using two bins are similar.}), while for our standard sample of field galaxies, the \cl\ method yields, for the same assumed $\to = 40$\,Myr, \CompCLs\,, \CompCLA\, and \PEforty. In contrast, previous production efficiencies derived for field galaxies were much lower; for instance, \citetalias{Maoz2012_fig} obtained $PE = 0.00130 \pm 0.00015\, \PEunit$ (reflecting the lower scale factor $A$ discussed above) and \citet{Maoz2017_DTD} derived $PE = 0.0013 \pm 0.0001\, \PEunit$.

Our higher production efficiency may influence conclusions about abundances. For instance, \citet{Maoz2017_DTD} have proposed that the spatial distribution of iron and $\alpha$-elements in the Milky Way (MW) may be explained by a simple model: the halo and thick disk consist of an old population, formed in a burst at $z\sim 3$, with the steeper DTD slope and higher SN~Ia production efficiency of cluster galaxies, while the thin disk would be explained by a more extended SFH, with the DTD derived for field galaxies. It would be interesting to see how a different production efficiency changes these results.

\subsection{DTD Scenarios}
\label{subsec:scenarios}

Next we consider which DTD scenario is favored using our slope and scale factor. A value of $s\sim -1.35$, assuming no DTD break, is intermediate between the generic predictions of the double-degenerate and the double detonation \citep{Ruiter2011_dtd} scenarios. This slope may also be possible for the SD channel with an appropriate distribution of binary mass ratios (see Fig. 7 of \citealt{Greggio2005_dtd} and the findings of \citealt{Moe2015_SD}), or if there exists a mechanism that delays the explosion. Therefore it is not possible to constrain the DTD channel from our analysis.

Furthermore, the DTD does not discriminate between scenarios if a break in the DTD is allowed: given the limitations of the models described in \S \ref{sec:methods}, only a part of the \slopes\ parameter space can be constrained by either the \cl\ or the \sfhr\ method (see Fig. \ref{Fig:method_comp}). Both methods seem to allow for a shallower slope at young ages, and a steeper slope at later times, although it is never steeper than $s_2=-2$ at the $68\%$ confidence level. Using our default sample and assuming $\tc=1$\,Gyr, we find \defss.

Finally, it is possible that more than one channel is realized in nature; in this case, the effective DTD need not have a simple form. For instance, \citet{Wang2013_V0} have suggested the possibility that high velocity (HV) SNe~Ia have a distinct progenitor different from that of normal velocity (NV) events; it would therefore be interesting to repeat the analysis of this paper separately for NV and HV events. 

\section{Summary}
\label{sec:summary}

We have studied the \cl\ and \sfhr\ methods for probing the DTD of SN~Ia. The \cl\ technique, first introduced by \citetalias{Heringer2017_DTD}, avoids the use of stellar masses and ages, two quantities that are notoriously difficult to measure. Furthermore, this method is insensitive to the SFH ($\it cf.$ the cluster method), performs the full Eq. \ref{eq:convolution} convolution to compute rates, and uses unbinned data ($\it cf.$ the \sfhr\ method).

In order for the \cl\ method to fit the normalization of the DTD, we improved the photometric corrections necessary to compute absolute magnitudes. These corrections were indirectly checked against an independent source, exhibiting excellent agreement. We showed that the \sfhr\ and the \cl\ methods lead to a consistent DTD slope, while the scale factors disagree at the $95\%$ confidence level (using the same dataset).

By applying the \cl\ method to our default dataset of field galaxies, and under the assumption of a continuous power-law slope, we obtained \defA\ and \defs. Unlike previous inferences for field galaxies, we find values that are in reasonable agreement with the production efficiency of \citealt{Friedmann2018_ClusterDTD} for galaxy clusters. A study of systematic uncertainties indicates that the slope $s$ and scale factor $\rm{log}\ A$ can be determined to within $\sim 0.1$ and $\sim 0.2\,$dex, respectively. Based on these results, we do not favor any particular DTD scenario.

There are several ways in which the measured DTD and its interpretation could be improved. A much larger sample of SNe could be obtained by relaxing the requirement that host galaxies have spectroscopy; this would improve the statistical errors in the broken power-law slope measurements, and provide an improved constraint on proposed DTD scenarios. A larger sample would furthermore allow a test of whether SN events with different photometric and spectroscopic properties have different DTD properties. A specific example of this is the prediction that NV and HV events form two distinct populations of SN~Ia (\citealt{Wang2013_V0}); with a large enough sample of NV and HV events, one could test whether these events are described by distinct DTD parameters and hence evolutionary scenarios.

\acknowledgments
We thank Dani Maoz for his insightful comments and for providing us with the \vespa\ data used in his original work. We also thank the organizers of the ``Observational Signatures of Type Ia Supernova Progenitors III'' workshop, which led us to improve the methods to measure the delay times of SNe~Ia.

\software{
Astropy \citep{Astropy2013},
KCORRECT (version~4.3~\citealt{blanton2007_kcorrect}),
FSPS (version~3.0~\citealt{Conroy2009_fspsI,Conroy2010_fspsII}}),
Python-FSPS (version~2017.07.05~\citealt{Mackey_PythonFSPS}).

\appendix


\section{VESPA Rates as a Function of DTD Parameters}
\label{sec:app_rates}

For completeness, here we expand Eq. \ref{eq:approx_rates} to a form where it depends only on the DTD parameters. The mean ages are set by the domain of the age bins in \citetalias{Maoz2012_fig}, except that we leave the unknown early age of the first bin as a free parameter; the mean ages are then

\begin{equation}
\bar{t}_1 = \frac{\to(\mathrm{Gyr}) + 0.42}{2}\, \mathrm{Gyr},\,\,\,\, \bar{t}_2 = \frac{0.42 + 2.4}{2}=1.41\, \mathrm{Gyr}\,\,\,\, \mathrm{and}\,\,\,\, \bar{t}_3 = \frac{2.4 + 14}{2}=8.2\, \mathrm{Gyr}. 
\label{eq:mean_ages}    
\end{equation}

\noindent Eq. \ref{eq:approx_rates} therefore becomes 

\begin{equation}
\begin{aligned}
r_{i,1}  &= A \times m_{i,1} \times \bar{t}_1 ^{s_1}, \\
r_{i,2}  &= A \times m_{i,2} \times \bar{t}_2 ^{s_1}\,\,\,\, \mathrm{if}\,\,\,\, \bar{t}_2 \leq \tc \,\,\,\, \\
r_{i,2} &= B \times m_{i,2} \times \bar{t}_2 ^{s_2}\,\,\,\, \mathrm{if}\,\,\,\, \bar{t}_2 \geq \tc, \\
r_{i,3}  &= B \times m_{i,3} \times \bar{t}_3 ^{s_2}.
\end{aligned}
\label{eq:approx_expanded}    
\end{equation}

\newpage

\section{Determining Most Likely Parameters}
\label{sec:app_pars}

Here we show in detail how our likelihoods are computed. This is similar to the implementation of \citet{Maoz2010_powerlaw}, \citetalias{Maoz2012_fig}, \citet{Gao2012_SNmatching} and \citetalias{Heringer2017_DTD}.

We first assign a rate to each galaxy (denoted by the subscript $i$) in the data set. This is accomplished by using the \sfhr\ or the \cl\ methods:

\begin{equation}
\begin{aligned}
r_i (A,s_1,s_2) =& \, \sum_j \mathcal{V}(m_{ij},\langle \mathrm{DTD}_{j}(A,s_1,s_2) \rangle) \,\, \mathrm{[\unity]}, \\
s_i (A,s_1,s_2) =& \, \mathcal{S}(\Delta_i,\mathrm{DTD}(A,s_1,s_2)) \,\,  \mathrm{[\unitS]},
\end{aligned}
\label{eq:def_rate}    
\end{equation}

\noindent where $j$ runs through each age bin ($j={1,2,3}$) in the \sfhr\ analysis; $\mathcal{V}$ denotes the application of the \sfhr\ relation from Eq. \ref{eq:M12_rates}, and $\mathcal{S}$ the application of the \relation\ relation, as established in Fig. \ref{Fig:sSNRL}. 

Next, these rates are converted to unitless absolute rates through the following relations:

\begin{equation}
\begin{aligned}
\lambda_i (A,s_1,s_2) =& \, \epsilon_i(z_i)\,t_i(z_i)\, r_i (A,s_1,s_2), \\
\lambda_i (A,s_1,s_2) =& \, L_{r,i} \epsilon_i(z_i)\,t_i(z_i)\, s_i (A,s_1,s_2),
\end{aligned}
\label{eq:abs_rate}    
\end{equation}

\noindent where $z_i$ is the galaxy redshift and $t_i$ corresponds to the approximate observing time window for each galaxy. As in \citetalias{Maoz2012_fig}, we adopt

\begin{equation}
t_i = \frac{269}{365.25\, (1 + z_i)}\,\, [\mathrm{yr}],
\label{eq:time_window}    
\end{equation}

\noindent while $\epsilon_i$ is a detection efficiency estimated by \citet{Dilday2010_SNrate} and parameterized by \citetalias{Maoz2012_fig} as

\begin{equation}
        \epsilon_i =
        \left\{ \begin{array}{ll}
            0.72 & \textrm{if}\,\,\, 0 \leq z_i \leq 0.175 \\
            -3.2\, z_i + 1.28 & \textrm{if}\,\,\, 0.175 \leq z_i \leq 0.4.
        \end{array} \right.
\label{eq:efficiency} 
\end{equation}

\noindent Note that \citetalias{Maoz2012_fig} also tried a slightly different detection efficiency function, which would start declining past $z=0.184$ ($z=0.166$) for star forming (passive) galaxies. We do not adopt these extra corrections because \textit{(i)} the values we use for comparison (from their Table 2) do not use this modified function, \textit{(ii)} these corrections are likely small for a sample trimmed at $z=0.2$, and \textit{(iii)} information regarding whether a galaxy is star forming or not is not always available.

To derive likelihoods, one can treat the SN~Ia rate problem as a Poisson experiment in which the expected number of events during the survey time is very small. Thus, following the approach of \citet{Cash1979_likelihood}, the probability of a galaxy hosting $n_i$ events given an expected rate $\lambda_i$ is

\begin{equation}
P_i(n_i|\lambda_i (A,s_1,s_2)) \approx \frac{\lambda_i ^{n_i}\, e^{-\lambda_i}}{n_i !}.
\label{eq:poisson}    
\end{equation}

The corresponding likelihood $L$ of a given DTD model will then be the simple multiplication over all galaxies of the probabilities in Eq. \ref{eq:poisson}: 

\begin{equation}
L (A,s_1,s_2) = \prod _i P_i(n_i|\lambda_i (A,s_1,s_2)),
\label{eq:likelihood1}    
\end{equation}

\noindent which can be simplified in logarithmic space to

\begin{equation}
\mathrm{ln}\, L (A,s_1,s_2) = \sum _i \mathrm{ln}\, \left( \frac{\lambda_i ^{n_i}\, e^{-\lambda_i}}{n_i !} \right).
\label{eq:likelihood2}    
\end{equation}

Because all galaxies in the sample have either $n_i = 0$ (non hosts) or $n_i = 1$ (hosts; numbered with the subscript $k$), Eq. \ref{eq:likelihood2} simplifies to:

\begin{equation}
\mathrm{ln}\, L (A,s_1,s_2) = - \sum _{i \neq k} \lambda_i + \sum _k \mathrm{ln}\, \left( \lambda_k \, e^{-\lambda_k} \right) = - \sum _{i \neq k} \lambda_i + \sum _k \mathrm{ln}\, \left(  \lambda_k \right) - \sum _k \lambda_k = - \sum _i \lambda_i + \sum _k \mathrm{ln}\, \left(  \lambda_k \right).
\label{eq:likelihood3}    
\end{equation}

\noindent Given the identity $N_{\rm{exp}} \equiv \sum _i \lambda_i$, Eq. \ref{eq:likelihood3} becomes

\begin{equation}
\mathrm{ln}\, L (A,s_1,s_2) = - N_{\mathrm{exp}} + \sum _k \mathrm{ln}\, \left(  \lambda_k \right).
\label{eq:likelihood4}    
\end{equation}

Next, we show that the computational cost of surveying the parameter space of ($A$,$s_1$,$s_2$) can be reduced by decoupling the scale factor $A$ and treating this parameter analytically. First, assume that the likelihood $L$ is computed for a scale factor $A_0(s_1,s_2)$, which ensures that the predicted number of events matches the observations, $N_{\rm{exp}} = N_{\rm{obs}}$. Therefore, from Eq. \ref{eq:likelihood4},

\begin{equation}
\mathrm{ln}\, L (A_0,s_1,s_2) = - N_{\mathrm{obs}} + \sum _k \mathrm{ln}\, \left(  \lambda_k (A_0,s_1,s_2) \right).
\label{eq:likelihood5}    
\end{equation}

\noindent Because the specific rates in Eq. \ref{eq:def_rate} are linear with respect to $A$, one can define $A \equiv f \times A_0$ so that the predicted absolute rates in Eq. \ref{eq:abs_rate} can be written as $\lambda_i (A,s_1,s_2) = f \times \lambda_i (A_0,s_1,s_2)$. Eq. \ref{eq:likelihood3} then becomes

\begin{equation}
\begin{split}
\mathrm{ln}\, L (A,s_1,s_2) &= - \sum_i f\, \lambda_i (A_0,s_1,s_2) + \sum _k \mathrm{ln}\, [  f \lambda_k (A_0,s_1,s_2) ] \\ &= - f  \sum_i \lambda_i (A_0,s_1,s_2) + \sum _k \mathrm{ln}\, \lambda_k (A_0,s_1,s_2)  + \sum _k \mathrm{ln} (f)  . \\&= 
-f N_{\mathrm{obs}} + [N_{\mathrm{obs}} + \mathrm{ln}\, L (A_0,s_1,s_2)] +  N_{\mathrm{obs}} ln(f),
\label{eq:likelihood6}
\end{split}
\end{equation}

\noindent which simplifies to

\begin{equation}
\mathrm{ln}\, L (A,s_1,s_2) = N_{\mathrm{obs}}\, [1 - f + ln(f)] + \mathrm{ln}\, L (A_0,s_1,s_2).
\label{eq:likelihood7}    
\end{equation}

\noindent Eq. \ref{eq:likelihood7} is well behaved and, for a given pair $s_1,s_2$, reaches a maximum when $f=1$ (i.e. $A=A_0$). This demonstrates the expected relation that the maximum likelihood occurs when $N_{\rm{exp}} = N_{\rm{obs}}$ and that marginalizing over the scale factor parameter would only correspond to a constant factor that does not depend on $s_1$ or $s_2$.

\newpage

\section{Mass Comparison}
\label{sec:app_masses}

Here we investigate whether the \cl\ method is able to produce reliable estimates for the scale factor $A$. For this purpose we need to ensure that our photometric corrections and absolute magnitudes are realistic. We do so in an indirect way, via mass comparisons with \vespa\ masses, and with masses obtained in an independent study \citep{Chang2015_mass}. Masses can be reported as ``present-day'' ($M_\ast$) or ``formed'' stellar masses ($M_{\rm{T}}$). Taking a SSP as an example, its present-day stellar mass will decrease with time, but its formed mass remains unchanged.

The \cl\ method does not directly assess galaxy masses to constrain the DTD parameters because the SFH itself is not fixed. However, given the computed absolute magnitude of a galaxy, a mass range can be estimated if an SFH is assumed. For this exercise, we select a subset of galaxies that belong to the red sequence by imposing $-0.1 < \Dcd < 0.1$ and by requiring \vespa\ masses in the young bins to be negligible ($m_{i,1} = m_{i,2} = 0$). We further assume that these objects may be described by an exponential SFH with a short timescale of $\tau = 1$\, Gyr. We assign an age of 10\, Gyr to these galaxies and estimate their masses via the mass-to-light ratio predicted by FSPS. A mass uncertainty is computed by assigning ages of 2.4 and 14\, Gyr, corresponding to the age range of the oldest mass bin in \citetalias{Maoz2012_fig}.

\begin{figure*}
\epsscale{1.15}
\plotone{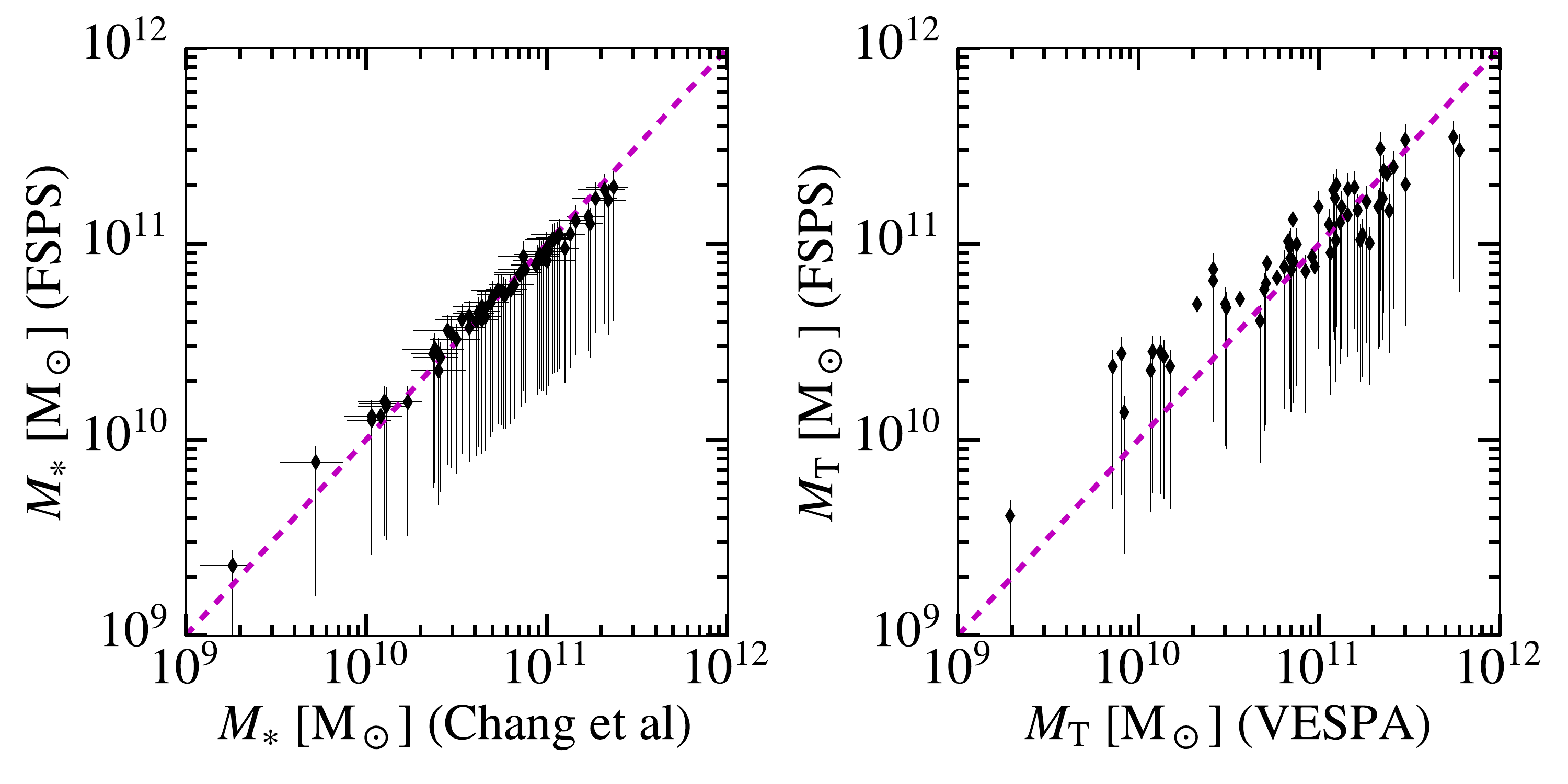}
\caption{Comparison between masses estimated from our model with those of \citet{Chang2015_mass} (left panel) and with \vespa\ masses (right panel). Masses from \citet{Chang2015_mass} are present-day stellar masses, and were obtained with the MAGPHYS package \citep{Cunha2008_MAGPHYS}, assuming a Chabrier IMF; their error bars are $\pm 2 \sigma$. \vespa\ masses (D. Maoz, private communication) represent formed masses, do not have uncertainties and were derived under the assumption of a Kroupa IMF. Our masses were computed using the FSPS package under the assumption of an exponential SFH with $\tau = 1$\, Gyr, spanning ages of 2.4--14 Gyr, and adopted an IMF that matched that of the works we compare against. Plotted are galaxies that belong to the red sequence. One can see a relatively good agreement of our estimated masses with those from Chang~{\it et al.}, thus giving us confidence in our photometric corrections.}
\label{Fig:mass}
\end{figure*}

In Fig. \ref{Fig:mass} we show the comparison between stellar masses from the \cl\ method and \citet{Chang2015_mass} (left panel), and the comparison between formed masses from \cl\ method and \vespa\ (right panel). For consistency, $\ssnrl$ models adopted a \citet{Chabrier2003_imf} IMF to compute present-day stellar masses and a \citet{Kroupa2001_imf} IMF to compute formed masses. Notwithstanding small biases, the agreement between our masses and those of \citet{Chang2015_mass} is excellent, and more than sufficient to validate our photometric corrections.

\newpage

\section{Data Set Comparison}
\label{sec:app_dataset}

\begin{figure*}
\epsscale{.85}
\plotone{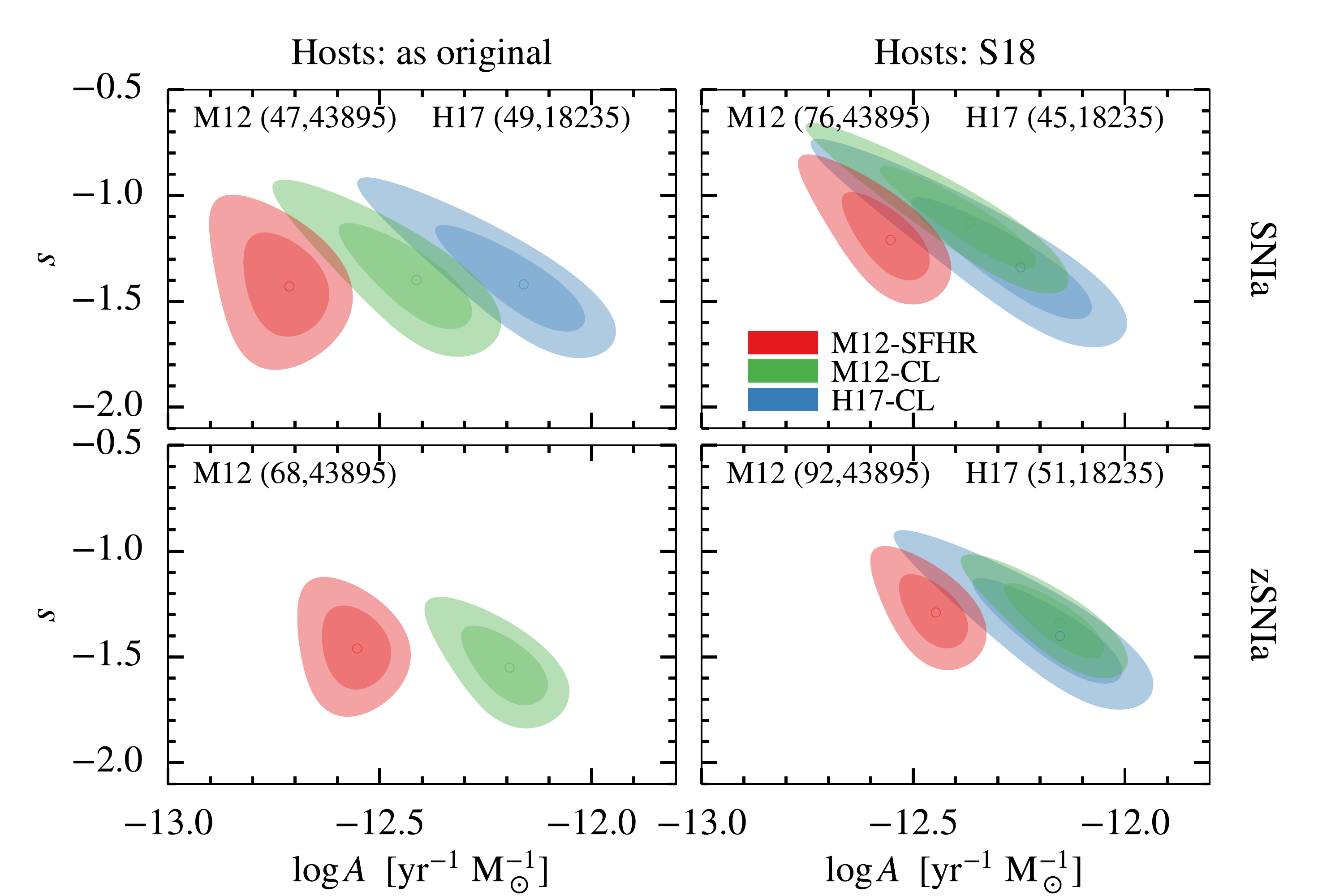}
\caption{DTD parameter likelihoods for different datasets. The confidence contours are computed using the \sfhr\ method applied to galaxies from the \citetalias{Maoz2012_fig} dataset (red), and using the \cl\ method applied to galaxies from the \citetalias{Maoz2012_fig} (green) and  \citetalias{Heringer2017_DTD} (blue) datasets. \textit{Left column:} hosts were taken from the original papers. \textit{Right column:} hosts were retrieved from \citetalias{Sako2018_SDSS}. \textit{Top row:} Only \citetalias{Sako2018_SDSS} class SNIa are included -- i.e. the SN~Ia were typed spectroscopically. \textit{Bottom row:} Both SNIa and zSNIa were included -- i.e. SN~Ia typed photometrically were also used, as long as the host had spectra. This shows that the most likely DTD parameters are significantly affected by the choice of dataset. For all datasets used here, we have imposed redshift ($0.01 < z < 0.2$) and photometry error cuts (see \S \ref{sec:data}) so that the \cl\ method could be applied. Numbers in brackets indicate the numbers of hosts and control galaxies. We have assumed $\to\ = 100\, \rm{Myr}$.}
\label{Fig:sample_comp}
\end{figure*}

In this appendix we compare the confidence contours calculated via the \cl\ and \sfhr\ methods for different datasets. This is shown in Fig. \ref{Fig:sample_comp}: the top panels use supernovae classified as SNIa, whereas the lower panels accept both SNIa and zSNIa classifications.  The left panels use hosts as defined in the original papers, whereas the right panels adopt the hosts of \citetalias{Sako2018_SDSS}. In order to apply the \cl\ method, we impose redshift and photometric error limits to all datasets (see \S \ref{subsec:data_thiswork}).  We note that the contours change little if $z_{\rm{max}}=0.4$ had been chosen instead of $z_{\rm{max}}=0.2$.

The slopes $s$ in Fig. \ref{Fig:sample_comp} are consistent, and therefore we focus on the scale factor. First, when both methods are applied to the same sample (red and green contours), the \sfhr\ method leads to smaller scale factors (as also noted in \S \ref{sec:results}). One possible explanation is that the \vespa\ masses tend to be higher than the estimated FSPS masses for the most massive galaxies in the sample. This trend is hinted at in the right panel of Fig. \ref{Fig:mass}, and we note that the effects of binning masses in the \sfhr\ method may suggest an even lower scale factor (see Fig. \ref{Fig:test_M12}). Understanding this discrepancy is beyond the scope of this work, but, as argued in appendix \ref{sec:app_masses}, we believe the scale factor derived from the \cl\ method is reliable.

Second, we have applied the same method (\cl) to control galaxies from \citetalias{Maoz2012_fig}'s and \citetalias{Heringer2017_DTD}'s datasets (green and blue contours, respectively). If the hosts are retrieved from the original works, then the scale factor assessed for \citetalias{Maoz2012_fig}'s dataset is much lower (top left panel). This makes sense, given that both datasets contain approximately the same number of hosts (in the chosen redshift range), but \citetalias{Maoz2012_fig}'s sample contains almost twice as many galaxies as \citetalias{Heringer2017_DTD}'s sample. This discrepancy is likely because there are some hosts missing in \citetalias{Maoz2012_fig}'s sample and is resolved when using hosts from \citetalias{Sako2018_SDSS}'s list.


\begin{thebibliography}{}
\expandafter\ifx\csname natexlab\endcsname\relax\def\natexlab#1{#1}\fi

\bibitem[{{Andreon} {et~al.}(2016){Andreon}, {Dong}, \&
  {Raichoor}}]{Andreon2016_cluster}
{Andreon}, S., {Dong}, H., \& {Raichoor}, A. 2016, \aap, 593, A2

\bibitem[{{Astropy Collaboration} {et~al.}(2013){Astropy Collaboration},
  {Robitaille}, {Tollerud}, {Greenfield}, {Droettboom}, {Bray}, {Aldcroft},
  {Davis}, {Ginsburg}, {Price-Whelan}, {Kerzendorf}, {Conley}, {Crighton},
  {Barbary}, {Muna}, {Ferguson}, {Grollier}, {Parikh}, {Nair}, {Unther},
  {Deil}, {Woillez}, {Conseil}, {Kramer}, {Turner}, {Singer}, {Fox}, {Weaver},
  {Zabalza}, {Edwards}, {Azalee Bostroem}, {Burke}, {Casey}, {Crawford},
  {Dencheva}, {Ely}, {Jenness}, {Labrie}, {Lim}, {Pierfederici}, {Pontzen},
  {Ptak}, {Refsdal}, {Servillat}, \& {Streicher}}]{Astropy2013}
{Astropy Collaboration}, {Robitaille}, T.~P., {Tollerud}, E.~J., {et~al.} 2013,
  \aap, 558, A33

\bibitem[{{Blanton} \& {Roweis}(2007)}]{blanton2007_kcorrect}
{Blanton}, M.~R., \& {Roweis}, S. 2007, \aj, 133, 734

\bibitem[{{Branch} {et~al.}(2009){Branch}, {Chau Dang}, \&
  {Baron}}]{Branch2009_subclasses}
{Branch}, D., {Chau Dang}, L., \& {Baron}, E. 2009, \pasp, 121, 238

\bibitem[{{Brandt} {et~al.}(2010){Brandt}, {Tojeiro}, {Aubourg}, {Heavens},
  {Jimenez}, \& {Strauss}}]{Brandt2010_dtd}
{Brandt}, T.~D., {Tojeiro}, R., {Aubourg}, {\'E}., {et~al.} 2010, \aj, 140, 804

\bibitem[{{Cash}(1979)}]{Cash1979_likelihood}
{Cash}, W. 1979, \apj, 228, 939

\bibitem[{{Chabrier}(2003)}]{Chabrier2003_imf}
{Chabrier}, G. 2003, \pasp, 115, 763

\bibitem[{{Chandrasekhar}(1931)}]{Chandrasekhar1931_mass}
{Chandrasekhar}, S. 1931, \apj, 74, 81

\bibitem[{{Chang} {et~al.}(2015){Chang}, {van der Wel}, {da Cunha}, \&
  {Rix}}]{Chang2015_mass}
{Chang}, Y.-Y., {van der Wel}, A., {da Cunha}, E., \& {Rix}, H.-W. 2015, \apjs,
  219, 8

\bibitem[{{Charlot} \& {Fall}(2000)}]{Charlot2000_dust}
{Charlot}, S., \& {Fall}, S.~M. 2000, \apj, 539, 718

\bibitem[{{Childress} {et~al.}(2015){Childress}, {Hillier}, {Seitenzahl},
  {Sullivan}, {Maguire}, {Taubenberger}, {Scalzo}, {Ruiter}, {Blagorodnova},
  {Camacho}, {Castillo}, {Elias-Rosa}, {Fraser}, {Gal-Yam}, {Graham}, {Howell},
  {Inserra}, {Jha}, {Kumar}, {Mazzali}, {McCully}, {Morales-Garoffolo},
  {Pandya}, {Polshaw}, {Schmidt}, {Smartt}, {Smith}, {Sollerman}, {Spyromilio},
  {Tucker}, {Valenti}, {Walton}, {Wolf}, {Yaron}, {Young}, {Yuan}, \&
  {Zhang}}]{Childress2015_Ni}
{Childress}, M.~J., {Hillier}, D.~J., {Seitenzahl}, I., {et~al.} 2015, ArXiv
  e-prints, arXiv:1507.02501

\bibitem[{{Conroy} {et~al.}(2009){Conroy}, {Gunn}, \&
  {White}}]{Conroy2009_fspsI}
{Conroy}, C., {Gunn}, J.~E., \& {White}, M. 2009, \apj, 699, 486

\bibitem[{{Conroy} {et~al.}(2010){Conroy}, {White}, \&
  {Gunn}}]{Conroy2010_fspsII}
{Conroy}, C., {White}, M., \& {Gunn}, J.~E. 2010, \apj, 708, 58

\bibitem[{{da Cunha} {et~al.}(2008){da Cunha}, {Charlot}, \&
  {Elbaz}}]{Cunha2008_MAGPHYS}
{da Cunha}, E., {Charlot}, S., \& {Elbaz}, D. 2008, \mnras, 388, 1595

\bibitem[{{Dilday} {et~al.}(2010){Dilday}, {Smith}, {Bassett}, {Becker},
  {Bender}, {Castander}, {Cinabro}, {Filippenko}, {Frieman}, {Galbany},
  {Garnavich}, {Goobar}, {Hopp}, {Ihara}, {Jha}, {Kessler}, {Lampeitl},
  {Marriner}, {Miquel}, {Moll{\'a}}, {Nichol}, {Nordin}, {Riess}, {Sako},
  {Schneider}, {Sollerman}, {Wheeler}, {{\"O}stman}, {Bizyaev}, {Brewington},
  {Malanushenko}, {Malanushenko}, {Oravetz}, {Pan}, {Simmons}, \&
  {Snedden}}]{Dilday2010_SNrate}
{Dilday}, B., {Smith}, M., {Bassett}, B., {et~al.} 2010, \apj, 713, 1026

\bibitem[{{Dotter}(2016)}]{Dotter2016_MIST}
{Dotter}, A. 2016, \apjs, 222, 8

\bibitem[{Foreman-Mackey {et~al.}(2014)Foreman-Mackey, Sick, \&
  Johnson}]{Mackey_PythonFSPS}
Foreman-Mackey, D., Sick, J., \& Johnson, B. 2014, python-fsps: Python bindings
  to FSPS (v0.1.1), doi:10.5281/zenodo.12157

\bibitem[{{Friedmann} \& {Maoz}(2018)}]{Friedmann2018_ClusterDTD}
{Friedmann}, M., \& {Maoz}, D. 2018, \mnras, 479, 3563

\bibitem[{{Frieman} {et~al.}(2008){Frieman}, {Bassett}, {Becker}, {Choi},
  {Cinabro}, {DeJongh}, {Depoy}, {Dilday}, {Doi}, {Garnavich}, {Hogan},
  {Holtzman}, {Im}, {Jha}, {Kessler}, {Konishi}, {Lampeitl}, {Marriner},
  {Marshall}, {McGinnis}, {Miknaitis}, {Nichol}, {Prieto}, {Riess}, {Richmond},
  {Romani}, {Sako}, {Schneider}, {Smith}, {Takanashi}, {Tokita}, {van der
  Heyden}, {Yasuda}, {Zheng}, {Adelman-McCarthy}, {Annis}, {Assef},
  {Barentine}, {Bender}, {Blandford}, {Boroski}, {Bremer}, {Brewington},
  {Collins}, {Crotts}, {Dembicky}, {Eastman}, {Edge}, {Edmondson}, {Elson},
  {Eyler}, {Filippenko}, {Foley}, {Frank}, {Goobar}, {Gueth}, {Gunn},
  {Harvanek}, {Hopp}, {Ihara}, {Ivezi{\'c}}, {Kahn}, {Kaplan}, {Kent},
  {Ketzeback}, {Kleinman}, {Kollatschny}, {Kron}, {Krzesi{\'n}ski}, {Lamenti},
  {Leloudas}, {Lin}, {Long}, {Lucey}, {Lupton}, {Malanushenko}, {Malanushenko},
  {McMillan}, {Mendez}, {Morgan}, {Morokuma}, {Nitta}, {Ostman}, {Pan},
  {Rockosi}, {Romer}, {Ruiz-Lapuente}, {Saurage}, {Schlesinger}, {Snedden},
  {Sollerman}, {Stoughton}, {Stritzinger}, {Subba Rao}, {Tucker}, {Vaisanen},
  {Watson}, {Watters}, {Wheeler}, {Yanny}, \& {York}}]{Frieman2008_SNsurvey}
{Frieman}, J.~A., {Bassett}, B., {Becker}, A., {et~al.} 2008, \aj, 135, 338

\bibitem[{{Frohmaier} {et~al.}(2019){Frohmaier}, {Sullivan}, {Nugent}, {Smith},
  {Dimitriadis}, {Bloom}, {Cenko}, {Kasliwal}, {Kulkarni}, {Maguire}, {Ofek},
  {Poznanski}, \& {Quimby}}]{Frohmaier2019_DTD}
{Frohmaier}, C., {Sullivan}, M., {Nugent}, P.~E., {et~al.} 2019, \mnras,
  arXiv:1903.08580

\bibitem[{{Gao} \& {Pritchet}(2013)}]{Gao2012_SNmatching}
{Gao}, Y., \& {Pritchet}, C. 2013, \aj, 145, 83

\bibitem[{{Graur} \& {Maoz}(2013)}]{Graur2013_fig}
{Graur}, O., \& {Maoz}, D. 2013, \mnras, 430, 1746

\bibitem[{{Graur} {et~al.}(2011){Graur}, {Poznanski}, {Maoz}, {Yasuda},
  {Totani}, {Fukugita}, {Filippenko}, {Foley}, {Silverman}, {Gal-Yam},
  {Horesh}, \& {Jannuzi}}]{Graur2011_dtd}
{Graur}, O., {Poznanski}, D., {Maoz}, D., {et~al.} 2011, \mnras, 417, 916

\bibitem[{{Graur} {et~al.}(2014){Graur}, {Rodney}, {Maoz}, {Riess}, {Jha},
  {Postman}, {Dahlen}, {Holoien}, {McCully}, {Patel}, {Strolger},
  {Ben{\'{\i}}tez}, {Coe}, {Jouvel}, {Medezinski}, {Molino}, {Nonino},
  {Bradley}, {Koekemoer}, {Balestra}, {Cenko}, {Clubb}, {Dickinson},
  {Filippenko}, {Frederiksen}, {Garnavich}, {Hjorth}, {Jones}, {Leibundgut},
  {Matheson}, {Mobasher}, {Rosati}, {Silverman}, {U}, {Jedruszczuk}, {Li},
  {Lin}, {Mirmelstein}, {Neustadt}, {Ovadia}, \& {Rogers}}]{Graur2014_fig}
{Graur}, O., {Rodney}, S.~A., {Maoz}, D., {et~al.} 2014, \apj, 783, 28

\bibitem[{{Greggio}(2005)}]{Greggio2005_dtd}
{Greggio}, L. 2005, \aap, 441, 1055

\bibitem[{{Heringer} {et~al.}(2017{\natexlab{a}}){Heringer}, {Pritchet},
  {Kezwer}, {Graham}, {Sand}, \& {Bildfell}}]{Heringer2017_DTD}
{Heringer}, E., {Pritchet}, C., {Kezwer}, J., {et~al.} 2017{\natexlab{a}},
  \apj, 834, 15

\bibitem[{{Heringer} {et~al.}(2017{\natexlab{b}}){Heringer}, {van Kerkwijk},
  {Sim}, \& {Kerzendorf}}]{Heringer2017_sequence}
{Heringer}, E., {van Kerkwijk}, M.~H., {Sim}, S.~A., \& {Kerzendorf}, W.~E.
  2017{\natexlab{b}}, \apj, 846, 15

\bibitem[{{Ivanova} {et~al.}(2013){Ivanova}, {Justham}, {Chen}, {De Marco},
  {Fryer}, {Gaburov}, {Ge}, {Glebbeek}, {Han}, {Li}, {Lu}, {Marsh},
  {Podsiadlowski}, {Potter}, {Soker}, {Taam}, {Tauris}, {van den Heuvel}, \&
  {Webbink}}]{Ivanova2013_CE}
{Ivanova}, N., {Justham}, S., {Chen}, X., {et~al.} 2013, \aapr, 21, 59

\bibitem[{{Komatsu} {et~al.}(2009){Komatsu}, {Dunkley}, {Nolta}, {Bennett},
  {Gold}, {Hinshaw}, {Jarosik}, {Larson}, {Limon}, {Page}, {Spergel},
  {Halpern}, {Hill}, {Kogut}, {Meyer}, {Tucker}, {Weiland}, {Wollack}, \&
  {Wright}}]{Komatsu2009_WMAP5}
{Komatsu}, E., {Dunkley}, J., {Nolta}, M.~R., {et~al.} 2009, \apjs, 180, 330

\bibitem[{{Kroupa}(2001)}]{Kroupa2001_imf}
{Kroupa}, P. 2001, \mnras, 322, 231

\bibitem[{{Kroupa}(2007)}]{Kroupa2007_imf}
---. 2007, ArXiv Astrophysics e-prints, astro-ph/0703124

\bibitem[{{Lejeune} {et~al.}(1997){Lejeune}, {Cuisinier}, \&
  {Buser}}]{Lejeune1997_BaSelI}
{Lejeune}, T., {Cuisinier}, F., \& {Buser}, R. 1997, VizieR Online Data
  Catalog, 412, 50229

\bibitem[{{Lejeune} {et~al.}(1998){Lejeune}, {Cuisinier}, \&
  {Buser}}]{Lejeune1998_BaSelII}
---. 1998, \aaps, 130, 65

\bibitem[{{Livio} \& {Mazzali}(2018)}]{Livio2018_review}
{Livio}, M., \& {Mazzali}, P. 2018, \physrep, 736, 1

\bibitem[{{Lupton} {et~al.}(1999){Lupton}, {Gunn}, \&
  {Szalay}}]{Lupton1999_system}
{Lupton}, R.~H., {Gunn}, J.~E., \& {Szalay}, A.~S. 1999, \aj, 118, 1406

\bibitem[{{Maoz} \& {Badenes}(2010)}]{Maoz2010_powerlaw}
{Maoz}, D., \& {Badenes}, C. 2010, \mnras, 407, 1314

\bibitem[{{Maoz} \& {Graur}(2017)}]{Maoz2017_DTD}
{Maoz}, D., \& {Graur}, O. 2017, \apj, 848, 25

\bibitem[{{Maoz} \& {Mannucci}(2012)}]{MaozandMannucci2012_dtd}
{Maoz}, D., \& {Mannucci}, F. 2012, PASA, 29, 447

\bibitem[{{Maoz} {et~al.}(2012){Maoz}, {Mannucci}, \& {Brandt}}]{Maoz2012_fig}
{Maoz}, D., {Mannucci}, F., \& {Brandt}, T.~D. 2012, \mnras, 426, 3282

\bibitem[{{Maoz} {et~al.}(2011){Maoz}, {Mannucci}, {Li}, {Filippenko}, {Della
  Valle}, \& {Panagia}}]{Maoz2011_fig}
{Maoz}, D., {Mannucci}, F., {Li}, W., {et~al.} 2011, \mnras, 412, 1508

\bibitem[{{Maoz} {et~al.}(2010){Maoz}, {Sharon}, \& {Gal-Yam}}]{Maoz2010_dtd}
{Maoz}, D., {Sharon}, K., \& {Gal-Yam}, A. 2010, \apj, 722, 1879

\bibitem[{{Marigo} \& {Girardi}(2007)}]{Marigo2007_Padova}
{Marigo}, P., \& {Girardi}, L. 2007, \aap, 469, 239

\bibitem[{{Marigo} {et~al.}(2008){Marigo}, {Girardi}, {Bressan}, {Groenewegen},
  {Silva}, \& {Granato}}]{Marigo2008_Padova}
{Marigo}, P., {Girardi}, L., {Bressan}, A., {et~al.} 2008, \aap, 482, 883

\bibitem[{{Moe} \& {Di Stefano}(2015)}]{Moe2015_SD}
{Moe}, M., \& {Di Stefano}, R. 2015, \apj, 801, 113

\bibitem[{{Niemack} {et~al.}(2009){Niemack}, {Jimenez}, {Verde}, {Menanteau},
  {Panter}, \& {Spergel}}]{Niemack2009_photoz}
{Niemack}, M.~D., {Jimenez}, R., {Verde}, L., {et~al.} 2009, \apj, 690, 89

\bibitem[{{Nomoto} {et~al.}(2000){Nomoto}, {Umeda}, {Kobayashi}, {Hachisu},
  {Kato}, \& {Tsujimoto}}]{Nomoto2000_SD_MSandRG}
{Nomoto}, K., {Umeda}, H., {Kobayashi}, C., {et~al.} 2000, in American
  Institute of Physics Conference Series, Vol. 522, Cosmic Explosions: Tenth
  Astrophysics Conference, ed. S.~S. {Holt} \& W.~W. {Zhang}, 35--52

\bibitem[{{Nugent} {et~al.}(1995){Nugent}, {Phillips}, {Baron}, {Branch}, \&
  {Hauschildt}}]{Nugent1995_sequence}
{Nugent}, P., {Phillips}, M., {Baron}, E., {Branch}, D., \& {Hauschildt}, P.
  1995, \apjl, 455, L147

\bibitem[{{Perrett} {et~al.}(2012){Perrett}, {Sullivan}, {Conley},
  {Gonz{\'a}lez-Gait{\'a}n}, {Carlberg}, {Fouchez}, {Ripoche}, {Neill},
  {Astier}, {Balam}, {Balland}, {Basa}, {Guy}, {Hardin}, {Hook}, {Howell},
  {Pain}, {Palanque-Delabrouille}, {Pritchet}, {Regnault}, {Rich},
  {Ruhlmann-Kleider}, {Baumont}, {Lidman}, {Perlmutter}, \&
  {Walker}}]{Perrett2012_DTD}
{Perrett}, K., {Sullivan}, M., {Conley}, A., {et~al.} 2012, \aj, 144, 59

\bibitem[{{Polin} {et~al.}(2018){Polin}, {Nugent}, \&
  {Kasen}}]{Polin2018_multiple}
{Polin}, A., {Nugent}, P., \& {Kasen}, D. 2018, arXiv e-prints,
  arXiv:1811.07127

\bibitem[{{Rodney} {et~al.}(2014){Rodney}, {Riess}, {Strolger}, {Dahlen},
  {Graur}, {Casertano}, {Dickinson}, {Ferguson}, {Garnavich}, {Hayden}, {Jha},
  {Jones}, {Kirshner}, {Koekemoer}, {McCully}, {Mobasher}, {Patel}, {Weiner},
  {Cenko}, {Clubb}, {Cooper}, {Filippenko}, {Frederiksen}, {Hjorth},
  {Leibundgut}, {Matheson}, {Nayyeri}, {Penner}, {Trump}, {Silverman}, {U},
  {Azalee Bostroem}, {Challis}, {Rajan}, {Wolff}, {Faber}, {Grogin}, \&
  {Kocevski}}]{Rodney2014_DTD}
{Rodney}, S.~A., {Riess}, A.~G., {Strolger}, L.-G., {et~al.} 2014, \aj, 148, 13

\bibitem[{{Ruiter} {et~al.}(2011){Ruiter}, {Belczynski}, {Sim}, {Hillebrandt},
  {Fryer}, {Fink}, \& {Kromer}}]{Ruiter2011_dtd}
{Ruiter}, A.~J., {Belczynski}, K., {Sim}, S.~A., {et~al.} 2011, \mnras, 417,
  408

\bibitem[{{Sako} {et~al.}(2011){Sako}, {Bassett}, {Connolly}, {Dilday},
  {Cambell}, {Frieman}, {Gladney}, {Kessler}, {Lampeitl}, {Marriner}, {Miquel},
  {Nichol}, {Schneider}, {Smith}, \& {Sollerman}}]{Sako2011_SNphoto}
{Sako}, M., {Bassett}, B., {Connolly}, B., {et~al.} 2011, \apj, 738, 162

\bibitem[{{Sako} {et~al.}(2018){Sako}, {Bassett}, {Becker}, {Brown},
  {Campbell}, {Wolf}, {Cinabro}, {D'Andrea}, {Dawson}, {DeJongh}, {Depoy},
  {Dilday}, {Doi}, {Filippenko}, {Fischer}, {Foley}, {Frieman}, {Galbany},
  {Garnavich}, {Goobar}, {Gupta}, {Hill}, {Hayden}, {Hlozek}, {Holtzman},
  {Hopp}, {Jha}, {Kessler}, {Kollatschny}, {Leloudas}, {Marriner}, {Marshall},
  {Miquel}, {Morokuma}, {Mosher}, {Nichol}, {Nordin}, {Olmstead}, {{\"O}stman},
  {Prieto}, {Richmond}, {Romani}, {Sollerman}, {Stritzinger}, {Schneider},
  {Smith}, {Wheeler}, {Yasuda}, \& {Zheng}}]{Sako2018_SDSS}
{Sako}, M., {Bassett}, B., {Becker}, A.~C., {et~al.} 2018, \pasp, 130, 064002

\bibitem[{{Salpeter}(1955)}]{Salpeter1955_imf}
{Salpeter}, E.~E. 1955, \apj, 121, 161

\bibitem[{{S{\'a}nchez-Bl{\'a}zquez} {et~al.}(2006){S{\'a}nchez-Bl{\'a}zquez},
  {Peletier}, {Jim{\'e}nez-Vicente}, {Cardiel}, {Cenarro},
  {Falc{\'o}n-Barroso}, {Gorgas}, {Selam}, \& {Vazdekis}}]{Sanchez2006_MILES}
{S{\'a}nchez-Bl{\'a}zquez}, P., {Peletier}, R.~F., {Jim{\'e}nez-Vicente}, J.,
  {et~al.} 2006, \mnras, 371, 703

\bibitem[{{Sand} {et~al.}(2012){Sand}, {Graham}, {Bildfell}, {Zaritsky},
  {Pritchet}, {Hoekstra}, {Just}, {Herbert-Fort}, {Sivanandam}, {Foley}, \&
  {Mahdavi}}]{Sand2012_MENeaCS_SNsurvey}
{Sand}, D.~J., {Graham}, M.~L., {Bildfell}, C., {et~al.} 2012, \apj, 746, 163

\bibitem[{{Schawinski}(2009)}]{Schawinski2009_sn}
{Schawinski}, K. 2009, \mnras, 397, 717

\bibitem[{{Strauss} {et~al.}(2002){Strauss}, {Weinberg}, {Lupton}, {Narayanan},
  {Annis}, {Bernardi}, {Blanton}, {Burles}, {Connolly}, {Dalcanton}, {Doi},
  {Eisenstein}, {Frieman}, {Fukugita}, {Gunn}, {Ivezi{\'c}}, {Kent}, {Kim},
  {Knapp}, {Kron}, {Munn}, {Newberg}, {Nichol}, {Okamura}, {Quinn}, {Richmond},
  {Schlegel}, {Shimasaku}, {SubbaRao}, {Szalay}, {Vanden Berk}, {Vogeley},
  {Yanny}, {Yasuda}, {York}, \& {Zehavi}}]{Strauss2002_SDSS}
{Strauss}, M.~A., {Weinberg}, D.~H., {Lupton}, R.~H., {et~al.} 2002, \aj, 124,
  1810

\bibitem[{{Sullivan} {et~al.}(2006){Sullivan}, {Le Borgne}, {Pritchet},
  {Hodsman}, {Neill}, {Howell}, {Carlberg}, {Astier}, {Aubourg}, {Balam},
  {Basa}, {Conley}, {Fabbro}, {Fouchez}, {Guy}, {Hook}, {Pain},
  {Palanque-Delabrouille}, {Perrett}, {Regnault}, {Rich}, {Taillet}, {Baumont},
  {Bronder}, {Ellis}, {Filiol}, {Lusset}, {Perlmutter}, {Ripoche}, \&
  {Tao}}]{Sullivan2006_SNrates}
{Sullivan}, M., {Le Borgne}, D., {Pritchet}, C.~J., {et~al.} 2006, \apj, 648,
  868

\bibitem[{{Tojeiro} {et~al.}(2007){Tojeiro}, {Heavens}, {Jimenez}, \&
  {Panter}}]{Tojeiro2007_VESPA}
{Tojeiro}, R., {Heavens}, A.~F., {Jimenez}, R., \& {Panter}, B. 2007, \mnras,
  381, 1252

\bibitem[{{Tojeiro} {et~al.}(2009){Tojeiro}, {Wilkins}, {Heavens}, {Panter}, \&
  {Jimenez}}]{Tojeiro2009_VESPA}
{Tojeiro}, R., {Wilkins}, S., {Heavens}, A.~F., {Panter}, B., \& {Jimenez}, R.
  2009, \apjs, 185, 1

\bibitem[{{Toonen} {et~al.}(2014){Toonen}, {Claeys}, {Mennekens}, \&
  {Ruiter}}]{Toonen2014_BPS}
{Toonen}, S., {Claeys}, J.~S.~W., {Mennekens}, N., \& {Ruiter}, A.~J. 2014,
  \aap, 562, A14

\bibitem[{{Totani} {et~al.}(2008){Totani}, {Morokuma}, {Oda}, {Doi}, \&
  {Yasuda}}]{Totani2008_powerlaw}
{Totani}, T., {Morokuma}, T., {Oda}, T., {Doi}, M., \& {Yasuda}, N. 2008,
  \pasj, 60, 1327

\bibitem[{{Tutukov} \& {Yungelson}(1981)}]{Tutokov1981_DD}
{Tutukov}, A.~V., \& {Yungelson}, L.~R. 1981, Nauchnye Informatsii, 49, 3

\bibitem[{{Wang} {et~al.}(2013){Wang}, {Wang}, {Filippenko}, {Zhang}, \&
  {Zhao}}]{Wang2013_V0}
{Wang}, X., {Wang}, L., {Filippenko}, A.~V., {Zhang}, T., \& {Zhao}, X. 2013,
  Science, 340, 170

\bibitem[{{Webbink}(1984)}]{Webbink1984_DD}
{Webbink}, R.~F. 1984, \apj, 277, 355

\bibitem[{{Westera} {et~al.}(2002){Westera}, {Lejeune}, {Buser}, {Cuisinier},
  \& {Bruzual}}]{Westera2002_BaSelIII}
{Westera}, P., {Lejeune}, T., {Buser}, R., {Cuisinier}, F., \& {Bruzual}, G.
  2002, \aap, 381, 524

\bibitem[{{Whelan} \& {Iben}(1973)}]{Whelan1973_SD}
{Whelan}, J., \& {Iben}, Jr., I. 1973, \apj, 186, 1007

\bibitem[{{Willmer}(2018)}]{Willmer2018_filters}
{Willmer}, C.~N.~A. 2018, \apjs, 236, 47

\bibitem[{{Wyder} {et~al.}(2007){Wyder}, {Martin}, {Schiminovich}, {Seibert},
  {Budav{\'a}ri}, {Treyer}, {Barlow}, {Forster}, {Friedman}, {Morrissey},
  {Neff}, {Small}, {Bianchi}, {Donas}, {Heckman}, {Lee}, {Madore}, {Milliard},
  {Rich}, {Szalay}, {Welsh}, \& {Yi}}]{Wyder2007_cmd}
{Wyder}, T.~K., {Martin}, D.~C., {Schiminovich}, D., {et~al.} 2007, \apjs, 173,
  293

\end{thebibliography}
\end{document}